\documentclass[letterpaper,hyper]{JHEP3}
\usepackage{epsfig}
\usepackage{amsmath, amsfonts,euscript,bbm}
\usepackage{ifthen}
\usepackage{graphicx}
\pdfoutput=1

\newcommand{\roughly}[1]{\mathrel{\raise.3ex\hbox{$#1$\kern-0.85em
\lower1ex\hbox{$\sim$}}}}

\def\be{\begin{equation}}
\def\beq\begin{equation}
\def\ee{\end{equation}}
\def\bea{\begin{eqnarray}}
\def\eea{\end{eqnarray}}
\def\NN{N_{\rm max}}
\def\NNsv{N_{\rm max}^{\rm sv}}
\def\NNmv{N_{\rm max}^{\rm mv}}
\def\nn{\nonumber}

\def\eqref#1{(\ref{#1})}

\def\UV{{\scriptscriptstyle U \kern-.1emV}}
\def\IR{{\scriptscriptstyle I\kern-.18em R}}

\newcommand{\tbx}{\tilde {\bf x}_i}
\newcommand{\Ket}[1]{\left|#1\right\rangle}

\newcommand{\Bra}[1]{\left\langle#1\right|}

\newcommand{\Expect}[1]{\left\langle #1 \right\rangle}
\newcommand{\Expectnoc}[1]{\left\langle #1 \right\rangle}
\newcommand{\Mpl}{M_{\mathrm{Pl}}}
\newcommand{\mb}[1]{\mathbf{#1}}

\def\dphi{\delta\phi}

\def\L{\Lambda}
\def\LL{\mathcal{L}}

\def\Vsr{V_{\mathrm{sr}}}
\def\G{\mathcal{G}(k_i)}
\def\F{\mathcal{F}(k_i)}
\def\R{\mathcal{R}}
\def\B{B_N}
\def\e{\epsilon_*}

\def\D{\mathcal{D}}
\def\a{\alpha}

\def\td{2\pi^2 \Delta_R^2}

\title{Resonant Trispectrum and a Dozen More Primordial $N$-point functions}

\author{Louis Leblond${}^{1}$, Enrico Pajer,${}^{2}$\\
$^1$ Perimeter Institute for Theoretical Physics,
  Waterloo, Ontario, Canada;\\
$^2$ Department of Physics, Cornell University, Ithaca, NY 14853\\
Emails: \email{lleblond@perimeterinsitute.ca, ep295@cornell.edu}}
\date{}

\abstract {We compute all $N$-point primordial curvature correlation functions from inflation at tree-level up to $N$ of order ten or more depending on the choice of parameters. This is achieved for resonant inflationary models in which the inflaton potential has a periodic modulation on top of a slow-roll flat term. These models find a natural UV completion in string theory implementation of axion monodromy.  Key to the success of our computation is the observation that gravitational interactions among the perturbations can be neglected, which we argue is justified for any model of single-field inflation with parametrically large non-Gaussianity. We provide a comprehensive review and detailed derivations of known consistency relations for squeezed and collinear limits, and generalize them to any $N$-point function.}
\preprint{pi-cosmo-199}
\begin{document}

\tableofcontents

\section{Introduction}
Despite the increasing precision of recent cosmological observations, the data are still compatible with Gaussian primordial perturbations as seeds of large scale structures. The most na\"ive models of inflation, namely slow-roll inflation, generate perturbations that are Gaussian to very high accuracy. Nevertheless, Gaussianity is not a generic prediction of inflation, and indeed several models are known that lead to some observable non-Gaussian signal. A future detection of primordial non-Gaussianity would provide a powerful discriminant between different models. Given the wealth of new information hidden in the shape of the various $N$-point correlation functions, non-Gaussianity also offers to precisely map the physics of inflation.

For a Gaussian power spectrum, all $N$-point connected correlation functions vanish. As a consequence, a simple way to measure non-Gaussianity is to look at higher point functions such as the bispectrum (three-points) or the trispectrum (four points). Given an inflationary model, it is often straightforward but computationally intensive to obtain these correlation functions. Most complications arise due to the coupling to gravity. For single field models of inflation, several bispectra have been computed (starting in earnest with \cite{Acquaviva:2002ud, Maldacena:2002vr}, see \cite{Bartolo:2004if, Koyama:2010xj} for reviews), together with a handful of trispectra \cite{Seery:2006vu, Huang:2006eha, Chen:2009bc, Seery:2008ax, Arroja:2009pd,Engel:2008fu} and some partial results for $N=5, 6$ \cite{Lin:2010ua}. No higher $N$-point function is known except in particular regions of momentum space\footnote{See \cite{Jarnhus:2007ia} and \cite{Leblond,Shandera:2008ai} for some comments about the general structure of odd versus even $N$-point functions in slow-roll inflationary models.}.

The present paper stems from the following intuitive observation: when the most important interactions among the perturbations come from the sector driving inflation (typically a scalar field) as opposed to the gravity sector, then the latter can be neglected in computing correlation functions. More concretely one can formulate a ``practical'' decoupling limit in which the metric is an unperturbed (homogeneous) inflationary background\footnote{Let us stress that this is different from usual choice of a flat gauge since in that case only the spatial components of the metric are unperturbed.} while the inflaton is expanded as $\phi(t) + \delta\phi(\vec{x}, t)$. Any correlation function for $\delta\phi$ can be computed and translated into a correlation function for the curvature perturbations $\R$ by performing a gauge transformation. This is very similar in spirit to \cite{Cheung}, where it was noticed that this decoupling is analogous to the equivalence theorem in particle physics: W and Z scattering amplitudes at energies above their mass can be computed using only the Goldstone (longitudinal) modes \cite{Cornwall}. A similar decoupling can also be obtained in the context of massive gravity \cite{ArkaniHamed:2002sp}. Our observation is already implicit in the current literature and we will not attempt to formalize this decoupling limit. Instead, we exploit it to compute quantities that would otherwise be inapproachable. More concretely, we focus on resonant inflationary models and we find that upon neglecting gravitational interactions the calculation simplifies to the point that we are able to compute all $N$-point curvature correlation functions at tree level for $N$ as large as ten or twenty depending on the parameters.  

Resonant models were first proposed in \cite{Chen:2008wn}, where it was noticed that in the presence of a periodic modulation in the inflationary potential, non-Gaussianity can be resonantly enhanced. These models found a natural UV completion\footnote{For another occurrence of oscillations in string theory models see \cite{Bean:2008na}.} in string theory constructions based on axion monodromy \cite{Silverstein:2008sg, McAllister:2008hb,Flauger:2009ab, Berg:2009tg}. The basic idea is to consider an axion that enjoys a shift symmetry to all orders in perturbation theory and use it as the inflaton \cite{Freese:1990rb}.
This symmetry can be broken by non-perturbative effects but also explicitly by some small effect. The latter possibility provides a mechanism to naturally protect the flatness of the inflaton potential in large field models of inflation. Since non-perturbative effects are generically present, one expects a potential of the form
\be\label{Vin}
V(\phi) = \Vsr(\phi)  + \L^4\cos\left(\frac{\phi}{f}\right)\,,
\ee
where $\phi$ is a canonically normalized axion with axion decay constant $f$, $\Lambda$ is some non-perturbatively generated scale and $\Vsr$ is an arbitrary slow-roll potential proportional to the explicit breaking of the shift symmetry. A potential of the form \eqref{Vin} has been obtained in some class of type IIB string theory compactifications \cite{McAllister:2008hb,Flauger:2009ab, Berg:2009tg}, where $\Vsr\simeq \mu \phi$. In the following we will not assume this specific linear dependence, but consider the more general case where \eqref{Vin} is monotonic, which translates into the inequality $b<1$ if we define $b\equiv \Lambda^4 /(f \Vsr')$. The classical evolution induced by \eqref{Vin} has two typical time scales: the Hubble parameter $H$ and $\omega\equiv |\dot\phi|/f$ whose ratio we denote by $\alpha\equiv\omega/H$.

In the regime $\alpha\gg1$, non-Gaussianity is greatly enhanced by a resonance between the perturbations and the background \cite{Chen:2008wn, Hannestad:2009yx,Flauger:2010ja}. Since non-Gaussianity is parametrically large in $\alpha$, the decoupling of gravity can be used and this drastically simplifies the calculation of correlation functions. In fact, we are able to compute the single-vertex contributions to any (connected) curvature correlation function $\Expect{\R^N}$, i.e.~for any $N$, at tree level and away from collinear singularities. Working in an expansion in small $b$ and large $\alpha$, we find
\bea \label{AN}
  \Expect{\R_{\mathbf{k}_1} \cdots\R_{\mathbf{k}_N}}^{\rm{single\;vertex}} 
  &=&  (2\pi)^3 \delta^3\left(\sum_i^N{\bf k_i}\right) A_N \B(k_i) \,, \\
 A_N &\equiv& (-)^N \frac{3 b \sqrt{2\pi}}{2} \alpha ^{2N-9/2}\Delta_R^{2N-2}(2\pi^2)^{N-1}\nonumber \,,\\
   \B(k_i)& \equiv &\frac{1}{ K^{N-3} \prod_i k_i^2} \left[ \sin\left(\frac{\phi_K}{f} \right)-\frac1\alpha \cos\left(\frac{\phi_K}{f} \right) \sum_{j,i} \frac{k_i}{k_j} + \mathcal{O}(\alpha^{-2})\right]\; ,\nn
 \eea
 where $K \equiv \sum k_i$, $\phi_K = \phi_* - \sqrt{2\e}\ln K/k_*$, $k_*$ is some pivot scale and $\phi_*$ the corresponding value of the scalar field. The power spectrum is well approximated by the slow roll result $\Delta_R^2 = \frac{H^2}{8\pi^2\epsilon}$ with $\epsilon$ the slow-roll parameter.  Notice that the shape only depends on the magnitude of the vectors but not on their respective orientation. This gives a total of N variables as opposed to the usual $3N-6$. Note also that the above result reproduces the bispectrum calculated in \cite{Flauger:2010ja} using the full perturbed metric.

We provide convincing evidence that the single-vertex diagram gives the largest contribution to any $N$-point function away from collinear limits up to some $\NN$ which depends on the parameters of the model (as summarized in figure \ref{constraints}), i.e.
\be \label{if}
 \Expect{\R_{\mathbf{k}_1} \cdots\R_{\mathbf{k}_N}}\Big|_{N<\NN} \simeq \Expect{\R_{\mathbf{k}_1} \cdots\R_{\mathbf{k}_N}}^{\rm{single\;vertex}} \,.
\ee
As figure of merit, for $\alpha=100$ and any $b<.1$, $\NN=9$, while for $\alpha=1000$ and any $b<.1$, $\NN=17$. For $N\geq\NN$ the number of multi-vertex diagrams is so large that collectively they become comparable with single-vertex diagrams, although they are individually much smaller. Computing all these diagrams is straightforward in principle, but becomes computationally intractable for large $N$.

One potential concern is the behavior of \eqref{if} in specific momentum configurations. Generically, an $N$-point function has singularities when one of the momenta (squeezed limit) or a combination thereof (collinear limit) is taken to zero. The squeezed limit is correctly captured by the subleading term in \eqref{AN} but collinear limits are not reproduced. Fortunately, relations are available to obtain the behavior of correlation functions near these singularities as first pointed out in \cite{Maldacena:2002vr} and subsequently studied and generalized in \cite{Creminelli:2004yq, Huang:2006eha,Cheung:2007sv,Seery:2008ax,Li:2008gg, Giddings:2010nc,Ganc:2010ff,RenauxPetel:2010ty}. We review the known consistency relations and generalize them to any $N$. We provide a detailed discussion of their derivation, both in momentum and coordinate space in appendix \ref{consistency}. Using these relations we can supplement \eqref{if} with the leading divergent terms near collinear singularities. As an example we explicitly write down the resonant trispectrum in section \ref{s:aN}.

The outline of this paper is as follows. In \S \ref{s:gs} we explain our general strategy of decoupling gravity. We then apply it to the resonant inflationary model which we introduce in \S \ref{s:c}. In \S \ref{s:sv}, we compute all single-vertex diagrams. We show that the first two leading terms in $\alpha$ are compatible with known results and that the consistency relation in the squeezed limit is obeyed for any $N$. We then estimate the size of multi-vertex diagrams by using consistency relations in the collinear limit in \S \ref{s:mv}. We combine all our results together and discuss  their limits of validity in \S\ref{s:aN}. We conclude in \S \ref{s:dc}. The appendices provide the Feynman rules for the theory under consideration and review and prove various useful consistency relations.


\section{The General Strategy of Decoupling Gravity}\label{s:gs}

In this section we outline the general strategy we will adopt in this paper for the computation of correlators of primordial perturbations. We are surely not the first to realize and use this strategy but we would like to make the statements that are often implicit in the literature as clear as possible. 

A typical theory of inflation may look like\footnote{We set the reduced Planck mass $M_{\rm Pl}^{\rm red}$ to one.}
\be 
 S= \int \sqrt{-g} R +\int \sqrt{-g} \LL_{\phi}\,,
\ee
with some Lagrangian $\LL_{\phi}$ describing the sector responsible for the energy momentum tensor that drives inflation, e.g.~a scalar field with a slow-roll flat potential.  
The standard procedure is to first find a quasi de Sitter solution of the classical equations of motions derived from $S$. 
One then expands the action in powers of the perturbations of the inflaton and the metric. In the presence of gravity this is a laborious computation due to the large number of interaction terms.
  
As we mentioned in the introduction, we observe that 
\begin{itemize}
    \item \emph{Single field models of inflation with parametrically large non-Gaussianity  admit a limit where the leading contribution to curvature correlation functions can be calculated using only the action for $\delta\phi$ while leaving the metric unperturbed}.
 \end{itemize}
By parametrically large, what we have in mind is that in the given model parameters exist that we can tune to make higher $N$-point functions arbitrarily large compared to the power spectrum which is held fixed. The basic idea is that gravitational interaction are naturally small in the context of inflation and therefore if there exists an independent parameter which allow large interactions it should be possible to neglect gravity. The end result is expected to be correct only to leading order in that parameter. 

The computational prescription is the following:
\begin{itemize}
   \item Solve the Friedmann and scalar field equations to find the classical background, say $\{a(t),\phi(t)\}$.
   \item Derive the interaction Hamiltonian $H_I$ for the scalar field perturbations $\dphi$ from the action $S=\int \sqrt{-g} \LL_{\phi}$, where $g$ is computed from the background, e.g.~$\sqrt{-g}=a(t)^3$.
      \item Compute the equal-time correlators $\Expect{\dphi(t)^N}$ in the in-in formalism using $H_I$ computed as above.
   \item Convert the correlators of scalar field perturbations computed at horizon exit into correlators of comoving curvature perturbations $\R$, which are frozen outside the horizon. This conversion can be done at linear order in perturbations, i.e. using $\dphi=\sqrt{2\e}\R$ (where $\e=-\dot H/H^2$ is the Hubble slow-roll parameter). Note that to ensure that the gauge transformation is linear one may have to perform it soon after horizon crossing. 
\end{itemize}
The resulting correlators $\Expect{\R^N}$ should be accurate at leading order in the parameter which makes non-Gaussianity large. 

Before we plead further, we would like to make the following disclaimer. The idea that correlators of perturbations might be dominated by interactions inherent to the inflationary sector and that gravity might give subleading corrections is certainly not new. Well known examples are given by models with small sound speed and/or derivative interactions such as DBI inflation\cite{DBI}, p-adic models \cite{padic}, ghost inflation \cite{Ghost}, quasi-single field models \cite{Chen:2009we} and galileon\cite{Burrage:2010cu}. In most cases, the authors computed the non-Gaussian signal following the prescription outlined above (see for example  \cite{Creminelli:2003iq, Gruzinov:2004jx,Barnaby:2010ke} and \cite{Huang:2006eha, Chen:2009bc, Arroja:2009pd,Cheung}).

The purpose of the present paper is not to give a general proof of the validity of the above strategy, but to use it to compute quantities, such as $N$-point primordial correlation functions for $N\gg4$, that are yet unknown and unthinkable to compute including gravitational interactions. The success of our computation provides additional motivations to both put the decoupling strategy on firmer ground, which we hope to come back to in the future, and to further employ the decoupling strategy to widen our knowledge of correlation functions from inflation.


\section{Background and Power Spectrum}\label{s:c}

In this section we start up with a review of the model and of the classical inflationary evolution that it generates. This constitutes the background for the computation of correlation functions. We then derive the spectrum and show that it agrees with previous results \cite{Flauger:2009ab,Flauger:2010ja}. For more details about the phenomenology of the spectrum and the bispectrum we refer the interested reader to \cite{McAllister:2008hb,Flauger:2009ab}.

Consider a modulated potential of the following form
\be\label{V}
V(\phi) = \Vsr(\phi)  + \L^4\cos(\phi/f)\,,
\ee
where $\Vsr(\phi)$ is a smooth potential that, by itself, would admit at least $60$ efoldings of slow-roll inflation and $\L$ and $f$ are two parameters with dimensions of mass. The sinusoidal term arises naturally in the presence of non-perturbative corrections to the inflaton potential, in which case $\L$ is some non-perturbatively generated scale. A familiar example in field theory is an axion that couples to some gauge sector with strength given by the axion decay constant $f$ \footnote{The consequences of this coupling for inflationary perturbations have been studied in \cite{Anber:2009ua} and more recently in \cite{Barnaby:2010vf}.}. Instanton configurations in the gauge sector can induce periodic corrections to the axion potential, which have periodicity $2\pi f$ for a canonically normalized field and $\L\propto e^{-1/g^2}$ with $g$ the gauge coupling. An axion then is a good inflaton candidate because it enjoys a shift symmetry to all orders in perturbation theory. Even when the shift symmetry is explicitly broken in a controlled way, e.g.~by the $\Vsr$ term in \eqref{V}, its remnants still protect the flatness of the potential against quantum corrections. This provides a technically natural realization of large field inflation. It has been argued \cite{McAllister:2008hb,Flauger:2009ab}  (see also \cite{Kaloper:2008fb}) that a certain class of string theory constructions leads to potentials of the form \eqref{V}. For the purpose of this paper we take the potential \eqref{V} as given and study its phenomenological consequences.

With a suitable $\Vsr$, which we assume, the potential \eqref{V} can drive a prolonged phase of exponential expansion of the universe. The classical homogeneous background resulting from solving the Einstein and scalar field equations at leading non-trivial order in $\L$ was computed in \cite{Flauger:2009ab}. It turns out to be useful to introduce the monotonicity parameter
\be \label{monp}
b \equiv \frac{\Lambda^4}{\Vsr'(\phi_*) f}\,.
\ee
For $b<1$ the potential \eqref{V} is monotonic (provided $\Vsr$ is, which we are assuming). Comparison with observational data\footnote{Strictly speaking this comparison have been carried out only for two explicit forms of $\Vsr$, linear \cite{Flauger:2009ab} and quadratic \cite{Pahud:2008ae}. On the other hand we expect qualitatively similar results for other slow-roll potentials $\Vsr$. For a partial list of constraints on oscillations in the CMB see \cite{list}.}, e.g.~with the power spectrum of the CMB, tells us that $b\ll 1$. This justifies working in an expansion in powers of $b$. Another restriction on our parameters is $f\ll\Mpl$. This choice is motivated by the fact that, despite many efforts, superPlanckian axion decay constants have not been realized in any string theory construction so far and there are general arguments \cite{Banks:2003sx} that they can not arise in controlled situations, i.e.~within the supergravity approximation with a perturbative string coupling. When $b\ll1$ and $f\ll \Mpl$, the sinusoidal oscillations in the potential \eqref{V} induce a time dependence of the background in addition to the one due to slow-roll inflation. We call $\omega$ the time frequency of the oscillations of the background, and we introduce the dimensionless parameter
\be \label{alpha}
\alpha\equiv \frac{\omega}{H}=\frac{|\dot \phi|}{H f}=\frac{\sqrt{2\e}}{f}\,,
\ee 
where $\e\equiv -\dot H/H^2$ is the first Hubble slow-roll parameter (evaluated at some pivot scale). Throughout this paper we will assume to be in the regime $\alpha\gg 1$, i.e.~that the time scale of the oscillations of the background is much shorter than the time scale of the slow-roll inflationary evolution. This is the regime in which cosmological perturbations are resonantly enhanced, as first noticed in \cite{Chen:2008wn}, by the interaction with the background. The study of this effect is the main focus of the present work.

The solution for the background at leading order in $b$ and for $\alpha\gg1$ is \cite{Flauger:2009ab,Flauger:2010ja}
\be \label{phit}
\phi(t)=\phi_0(t)-\frac{3bf^2}{\sqrt{2\e}}\sin\left(\frac{\phi_0(t)}{f}\right)\,,
\ee
where $\phi_0(t)$ is the time evolution of the scalar field $\phi$ without the sinusoidal modulation, which of course depends on the specific choice of $\Vsr(\phi)$. For our purposes we do not need to specify $\Vsr(\phi)$, so our results are valid for any slow-roll flat potential $\Vsr$. 
In the following it will be useful to rewrite \eqref{phit} as
\bea
\phi(X) &= &\phi_K + \sqrt{2\e} \ln X - \frac{3b f^2}{\sqrt{2\e}}\sin\left(\frac{\phi_K + \sqrt{2\e}\ln X}{f}\right)\,,\nonumber\\ 
\phi_K &= &\phi_* - \sqrt{2\e}\ln K/k_*\,,
\eea
where time is now parameterized in terms of $X= -K\tau$, with $ad\tau=dt$ being the conformal time, $K$ is arbitrary (it cancels out) and $k_*$ is some pivot scale relevant for the experiments under consideration.

\subsection{The Two-point Function}\label{s:2p}

In this section we compute the two-point function of primordial curvature perturbations following the general strategy outlined in section \ref{s:gs}, i.e.~using only the scalar field sector of the theory. We show that at leading order in $b\ll1$ and $\alpha\gg1$ this agrees with the computation performed in \cite{Flauger:2009ab,Flauger:2010ja}, where gravity was taken into account. 

Before proceeding, let us review the definitions of the degrees of freedom we will work with and our gauge choices. Generic scalar metric fluctuations, to first order in perturbation theory, can be written as
\be \label{g}
ds^2=-(1+2\Phi)dt^2+2a(t)^2B_{,i}dx^i dt+a(t)^2\left[(1-2\Psi)\delta_{ij}+2E_{,ij}\right]dx^idx^j\,,
\ee
where $a(t)$ is the scale factor and $\{\Phi,B,\Psi,E\}$ are scalar perturbations. At first order in the perturbations, after imposing the constraint equations and in the absence of anisotropic stress, the five perturbations (four in \eqref{g} plus $\dphi$) reduce to just two, that combine together into
\be
\R\equiv -\Psi-\frac{H}{\dot \phi}\dphi\,,
\ee
which is invariant under time diffeomorphisms. In comoving gauge the only degree of freedom is in the comoving curvature perturbations
\be \label{metric}
\dphi=0\,,\quad g_{ij}= a^2 e^{2 \R} \delta_{ij}\,.
\ee
With an appropriate time diffeomorphism one can go from comoving (or uniform density) gauge into spatially flat gauge where $\Psi=0$ and vice versa. Since $\R$ is gauge invariant the two gauges are related by $\sqrt{2\e}\Psi=-\dphi$ where $\sqrt{2\e}= |\dot\phi|/H$.\footnote{In this paper we assume that $\dot\phi < 0$. If the opposite is true, the sign of the gauge transformation will differ.}

Let us now consider the linearized classical equation of motion for the scalar field perturbations in momentum space $\dphi_k$ (due to rotational invariance of the background $\dphi$ depends only on the absolute value $k$ of $\mathbf{k}$) in spatially flat gauge:
\be \label{eom}
\ddot \dphi_k+\frac{k^2}{a^2}\dphi_k+3H\dot\dphi_k-3bH^2\alpha \dphi_k \cos \left(\frac{\phi_0}{f}\right)=0\,,
\ee
where the slow-roll part of the potential has been neglected. Going to comoving gauge, after a long but straightforward calculation one can rewrite the equation above as 
\be
\partial_x^2\R_k-\frac{2 (1+\delta+\e)}{x}\partial_x\R_k+\R_k\left[1+\frac{\partial_x^2\sqrt{2\e}}{\sqrt{\e}}-\frac{3b}{x^2}\alpha\cos\left(\frac{\phi_0}{f}\right)\right]=0 \,,
\ee
where $x\equiv - k \tau$ and $\delta$ is a Hubble slow-roll parameter\footnote{A summary of various definitions of slow-roll parameters and their relations can be found in the appendix of \cite{Flauger:2010ja}.} defined by $\delta\equiv \ddot H/(2H\dot H)$.  Because of the cosine part of the potential we have the following hierarchy of slow-roll parameters $\e\ll\delta\ll\dot\delta/H$. Now notice that
\bea 
\frac{(\sqrt{2\e})_{xx}}{\sqrt{\e}}&=&\frac1{x^2}\left[\frac{\dot\delta}{H}+\delta(1+\delta+3\e)+\e(1+2\e)\right]\simeq \frac{\dot\delta}{x^2H}\,,
\eea
In the background induced by the potential \eqref{V}, one finds \cite{Flauger:2009ab}
\bea
\frac{\dot\delta}{H}&\simeq &3 b \alpha\cos\left(\frac{\phi_0}{f}\right)\,.
\eea
Hence \eqref{eom} agrees with eq (2.17) in \cite{Flauger:2010ja}
\be\label{MS}
\R_{xx}-\frac{2 (1+\delta)}{x}\R_x+\R=0 \,,
\ee
at leading non-trivial order, i.e.~for $\e\ll\delta\ll\dot\delta/H$.

To summarize, we have substituted the relation $\delta \phi=\sqrt{2\e} \R$ into the equation of motion for $\phi$ and showed that the resulting equation is the same as the one used in \cite{Flauger:2010ja,Flauger:2009ab} at leading order. Since the power spectrum follows directly from the solution of the Mukhanov-Sasaki equation \eqref{MS}, we conclude that using just the $\phi$-sector and neglecting the interactions coming from gravity is enough to correctly reproduce the power spectrum with its oscillatory features, to leading order. 
 
The quantization of the perturbations proceeds as usual and we briefly sketch it in order to introduce our notation. 
The momentum space perturbations are related to the real space perturbations by
\be 
\dphi(\textbf{x},t)=\int \frac{d^3\textbf{k}}{(2\pi)^3}\dphi(\textbf{k},t)e^{i\textbf{k}\cdot \textbf{x}}\,,
\ee
where the three-vectors $\{\textbf{x},\textbf{k}\}$ are the comoving coordinates and momentum, respectively. 
$\dphi(\textbf{k},t)$ is determined by the solution $\dphi_k(t)$ of \eqref{eom} with Bunch-Davies initial conditions. Then the momentum space quantum field is given by
\bea
\dphi(\textbf{k},t) &=& \dphi_k(t) a_{\textbf{k}}+\dphi_k^*(t) a^\dagger_{-\textbf{k}}\nonumber\,,
\eea
where the creation and annihilation operators satisfy the commutation relation
\bea
\left[a_{\mathbf{k}}, a^\dagger_{-\mathbf{k}^\prime}\right] & = &(2\pi)^3\delta^3(\mathbf{k} + \mathbf{k}^\prime)\,.\nonumber
\eea
Parameterizing time with $x=-k\tau$ one finds 
\bea\label{mf}
\dphi_{k}(x)&=& \frac{H}{\sqrt{2k^3}}\sqrt{\frac{\pi}{2}}x^{3/2}H^{(1)}_{3/2}\nonumber\\
&= &\frac{-i H}{\sqrt{2k^3}} (1 - ix)e^{ix} + \mathcal{O}(\e, \delta,b)\,.
\eea
Note that the mode functions $\dphi_k$ are given by the standard de Sitter result plus two types of subleading corrections. First, slow-roll suppressed terms that would change the Hankel function into $H^{(1)}_{\nu}$ with $\nu=3/2+2\e+\delta$. These terms would correct our results for non-Gaussianity by a small logarithmic running so we neglect them from now on. If needed it is straightforward to keep these terms. Second, $b$ suppressed terms coming from the oscillation of the background. These terms are important to compute the leading correction to the slow-roll power spectrum, which turns out to be
 \bea\label{2pt}
 \Expectnoc{\R_\mb{k}\R_\mb{k'}} & = & (2\pi)^3 \delta^3(\mb{k}+\mb{k'}) |R_k|^2\nonumber\\
 |R_k|^2 & = & 2\pi^2 \frac{\Delta^2_\R}{k^3} \left[ 1+ 3b\left(\frac{2\pi}{\alpha}\right)^{1/2}\cos\left(\frac{\phi_k}{f}\right)\right]\,,
  \eea
where $\Delta_R^2 = \frac{H^2}{8\pi^2 \epsilon}$. On the other hand, as discussed in \cite{Flauger:2009ab,Flauger:2010ja}, for the bispectrum and higher-point correlation functions these terms can be neglected. In fact, the leading term at linear order in $b$ in the correlators arises when taking the $b^0$ terms from the mode functions and the $b$ term from $H_I$ (which of course does not appear in the computation of the spectrum) for the mode functions.


\section{Single-vertex Diagrams}\label{s:sv}

The $N$-point correlation functions of $\dphi_{\mathbf{k}}$ can be computed using the in-in formalism. We will perform the calculation using the canonical formalism, but we will organize it according to the structure of (in-in) Feynman diagrams following the notation and Feynman rules in \cite{vMS}.  In this section we compute the contribution to $N$-point scalar correlation functions coming from diagrams that have a single interaction vertex or equivalently no internal propagators. As an example, the corresponding Feynman diagrams for $N=3,4$ are depicted  in figure \ref{contact} (the relevant Feynman rules are collected in Appendix \ref{Feynman}). Multi-vertex diagrams will be discussed in the next section. We follow the strategy outlined in section \ref{s:gs}, i.e.~we only use the scalar field Lagrangian. 
 
For our canonically normalized field $\dphi$ the interactions comes purely from the potential, hence the Hamiltonian is 
\be
H_I(\tau)\equiv\int d^3x a^3 \sum_{N=3}^{\infty} \frac{V^{(N)}}{N!}\dphi(x)^N\,.
\ee
Since we are interested in tree-level single-vertex diagrams we start from the leading order result in perturbation theory 
\be\label{start}
\Expect{ \prod_{i=1}^N\dphi_{\mathbf{k}_i}(\tau) } = i \int^\tau_{-\infty} \frac{d\tau^\prime}{\tau^\prime H}\Expect{[\dphi_{\mathbf{k}_1}\cdots\dphi_{\mathbf{k}_N}(\tau), H_I(\tau^\prime)]}\,.
\ee
Here we have assumed a quasi de Sitter background where, to leading order in the slow-roll parameters\footnote{We are allowed to use the slow-roll approximation for the background even in the presence of oscillations because we are working at leading order in $b$. The largest contribution to the correlator comes from taking the linear in $b$ term in  $H_I$ while the $b^0$ terms from all other factors.}, the scale factor is given by $a(\tau) = - \frac{1}{\tau H}$. 
 \begin{figure} [ht]
\centering
\includegraphics[width=0.7\textwidth, angle=0]{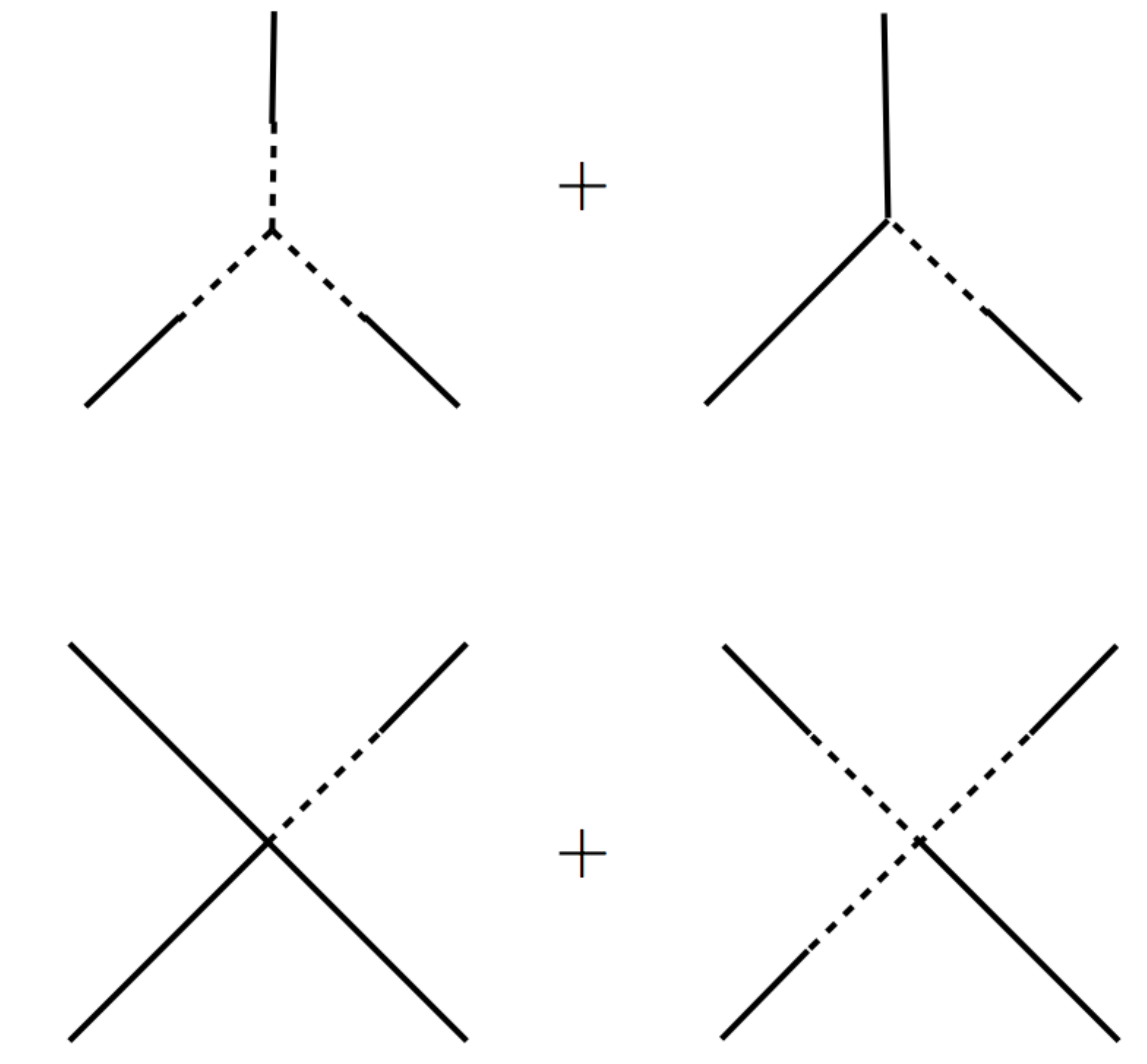}
\caption{Single-vertex Feynman diagrams for three- and four-point functions.  See Appendix \protect \ref{Feynman} for the Feynman rules. At $N=5$, there are three diagrams and so on.}\label{contact}
\end{figure}
Note that in order to simplify the notation, in \eqref{start} we have denoted the connected part of the N-point function by $\Expect{}$ instead of $\Expect{}_{c}$. In this paper we only refer to connected correlation functions so we always drop the label since no confusion can arise. To compute the commutator in \eqref{start}, we expand the real-space fields $\dphi(x)$ appearing in $H_I$ in Fourier modes. Since it is $\R$ and not $\dphi$ that is constant outside the horizon, it is customary to evaluate $\Expect{\dphi(\tau)^N}$ at the time of horizon crossing and perform a gauge transformation at that time to the variable $\R$.  This is not crucial since the evolution of $\dphi$ is a small slow-roll suppressed effect, which should precisely cancels the time evolution of the factor $\sqrt{2\e}$ involved in the change of gauge. As compared to the effect of oscillations, these effects together with other slow-roll suppressed corrections are subleading and we neglect them throughout our computation. Hence, it is consistent to focus on the limit $\tau \rightarrow 0$ where the relevant commutator is
\be
\Expect{ \left[\prod_{i=1}^N\dphi_{\mathbf{k}_i}(0),\prod_{j=1}^N \dphi_{ \mathbf{\tilde k}_j}^N(\tau')\right] }= N! \frac{H^{2N}}{\prod_i 2 k_i^3} (2\pi)^{3N}\left[e^{iK\tau^\prime}\prod_{i=1}^N (1-ik_i\tau^\prime) \delta^3(\mathbf{k}_i-\mathbf{\tilde k}_i) - \rm{c.c}\right]\,,
\ee
where $K =\sum_i k_i$ and we have subtracted the divergences by normal ordering. Putting this back in (\ref{start}) and performing the $d^3\tilde k_i$ and $d^3x$ integrals gives
\bea\label{ts}
\Expect{ \prod_{i=1}^N\dphi_{\mathbf{k}_i}(0) }  & = &\frac{(2\pi)^3 \delta^3(\sum_i \mathbf{k}_i)  H^{2N - 4}}{\prod_i 2 k_i^3}  I_{N}\,,\\
I_{N} & = &\int^ {\tau\rightarrow 0}_{-\infty} \frac{d\tau^\prime}{(\tau^\prime)^4} (-i) V^{(N)}\left[ (1-ik_1\tau^\prime)\cdots(1-ik_N\tau^\prime)e^{iK\tau^\prime} - \rm{c.c}\right]\,.\label{I}
\eea
Now normally $V^{(N)}$ is a constant and can be taken out of the integral in which cases one can show that \cite{Zaldarriaga:2003my}
\be
\frac{I_N}{V^{(N)}} \approx -\frac23 \sum k_i^3 N_e\; ,
\ee
where $N_e = \ln(-K \tau)$ is the number of efoldings. For slow-roll models, the contribution from this part of the Hamiltonian is actually slow-roll suppressed compared to the coupling to gravity so this is usually dropped (the $N_e$ enhancement  disappears when converted to $\R$ in single field models of inflation).  For resonant models, this term gives a large (often leading) contribution to $N$-point correlation functions as we will show.

In order to compute the integral \eqref{ts} we have to specify $V^{(N)}$. Let us consider the potential in \eqref{V}. The slow-roll\footnote{By "slow-roll" we mean not only the condition $\epsilon,\delta\ll1$, but also that all higher slow-roll parameters (e.g. see \cite{Liddle:1994dx}) are much smaller than one. In the presence of sinusoidal oscillations the first condition is satisfied while the second is not.} term $\Vsr$ gives rise to interactions much weaker than those coming from the the oscillatory part and can hence be neglected. Therefore, to leading order in the slow-roll approximation, we consider
\be
V(\phi) = 3\alpha b f^2H^2\cos(\phi/f)\,,
\ee
where the monotonicity parameter $b$ was defined in \eqref{monp} and $\alpha$ in \eqref{alpha}. Expanding this potential in perturbations we find the following $N$-th derivatives
\be\label{VN}
V^{(N)} = \frac{3\alpha b H^2}{f^{N-2}} \left\{\begin{array}{cc} (-)^{N/2} \cos(\phi/f) & , \; N \; \rm{even}\\
(-)^{\frac{N+1}{2}}\sin(\phi/f) & , \; N \; \rm{odd}\,, \end{array}\right.
\ee
In the above expression the time dependence is in $H$ and $\phi$. On the inflationary background we can use $\phi(X)/f=\phi_K/f+\alpha \log X$, where we parametrize time with $X \equiv -K\tau$ and $K$ is arbitrary at this point. After expanding out the products in \eqref{I}, we find integrals of the schematic form
\be
\int dX X^n \sin\left(X+\theta_1\right)\sin\left(\alpha\log X+\theta_2\right)\,,
\ee
for some integer $n$ and some real phases $\theta_1$ and $\theta_2$. The phase of the integral is stationary at $\tilde X=\alpha$ and this is where the integral gets its main contribution. These kind of integrals can be performed using the stationary phase approximation\footnote{In fact, a direct integration is possible as well and gives some combinations of $\Gamma$ functions. On the other hand, at the order we are working these $\Gamma$ functions should be expanded using Stirling's formula such that the end result is the same as the one from the stationary phase approximation.}
\bea \label{statphase}
\lim_{k\rightarrow+\infty}\int_{-\infty}^{\infty}G(x) e^{i kf(x)} dx&=&G(\tilde x)e^{ikf(\tilde x)} \nonumber\\
&&\hspace{-5cm}\times \left[\theta(f''(\tilde x))\sqrt{\frac{2\pi}{f''(\tilde x)}}e^{i\pi/4}+\theta(-f''(\tilde x))\sqrt{\frac{2\pi}{|f''(\tilde x)|}}e^{-i\pi/4}\right]\,,
\eea
where $\tilde x$ is the point where the phase $f(x)$ becomes stationary, i.e. $f(\tilde x)'=0$.
For example using this formula one can compute 
\bea\label{integral}
\int dX f(X) \cos(a + \alpha \ln X) \sin (X) &\sim &-f(\tilde X) \frac{\sqrt{2\pi}}{2} \alpha^{1/2} \sin(a-\theta)\nonumber\,,\\
 \int dX f(X) \sin(a + \alpha \ln X) \cos(X) & \sim&  f(\tilde X) \frac{\sqrt{2\pi}}{2} \alpha^{1/2} \sin(a-\theta)\nonumber\,,\\
  \int dX f(X) \cos(a + \alpha \ln X) \cos(X) & \sim& f(\tilde X) \frac{\sqrt{2\pi}}{2} \alpha^{1/2} \cos(a-\theta )\nonumber\,,\\
    \int dX f(X) \sin(a + \alpha \ln X) \sin(X) & \sim& f(\tilde  X) \frac{\sqrt{2\pi}}{2} \alpha^{1/2} \cos(a-\theta )\,,
\eea
where the phase is stationary at $X=\tilde X= \alpha $ and $\theta$ is a fixed (irrelevant) phase. We will have a sizable effect if $\alpha\gg1$, which physically corresponds to the regime when the time scale of the oscillation of the background is much shorter than the time scale of the slow-roll evolution. In this regime, the leading term is the one with the highest power of $X$.

\subsection{The Leading Contribution}\label{s:lead}
The whole integral \eqref{ts} can be computed (leading to \eqref{allsub}), but here we proceed step by step and first compute the leading piece. The leading term in powers of $\alpha$ comes from the highest power of $X$ in the integrand of \eqref{ts}
\be\label{X}
I_N = -i \int_0^\infty dX \; V^{(N)}(X) X^{N-4}\G \left[[(i)^N e^{-iX} - \rm{c.c} \right] +\dots\,,
\ee
where the dots stand for terms suppressed by at least one power of $1/\alpha$ and we have introduced the notation
\be 
\G = \frac{\prod_i k_i}{K^{N-3}}\,.
\ee 
Plugging the explicit expression for $V^{(N)}$ \eqref{VN} into $I_N$ as given in \eqref{X}, we obtain
\be
I_{N} = -2 \frac{3\alpha b H^2}{f^{N-2}} \G   \int_0^\infty dX X^{N-4} \left\{\begin{array}{cc}  \sin X \cos\left(\frac{\phi_k}{f}+\alpha\ln X \right)& , \; N \; \rm{even}\\
\cos X \sin\left(\frac{\phi_k}{f}+\alpha\ln X\right) & , \; N \; \rm{odd} \end{array}\right. \; .
\ee
Using \eqref{integral}, we find that both integrals give the same answer up to a sign. Modulo a $k$-independent irrelevant phase, the result is
\be
I_{N} = (-)^N \sqrt{2\pi \alpha} \frac{3\alpha b H^2}{f^{N-2}} \G \alpha ^{N-4} \sin \left(\frac{\phi_k}{f}\right) \label{lead}\; .
\ee
We can now write the leading contribution to the curvature correlation functions by performing the gauge transformation $\sqrt{2\e} \R = \delta\phi$. We find
\be
\Expect{ \prod_{i=1}^N\R_{\mathbf{k}_i} }  = (2\pi)^3 \delta^3\left(\sum_{i=1}^N \mathbf{k}_i\right)  \frac{A_N}{ K^{N-3} \prod_i k_i^2} \sin\left(\frac{\phi_K}{f} \right)\,,\label{leadR}
\ee
 where
 \bea\label{due}
 A_N &\equiv& \label{tre}(-)^N \frac{3 b \sqrt{2\pi}}{2} \alpha ^{2N-9/2}(2\pi^2\Delta_R^{2})^{N-1}\,, 
 \eea
 with $\Delta_R^2 = \frac{H^2}{8\pi^2\e}$. It is straightforward to check that for $N=3$ this result agrees at leading order with the one obtained in \cite{Flauger:2010ja}, where also the interactions induced by gravity were taken into account. 


\subsection{Main Result and Subleading Contributions}\label{ss:subl}

It is also interesting to compute the first subleading terms, i.e.~those with the second largest power of $\tau'$ (or equivalently of $X$) in \eqref{ts} since they are formally divergent in some squeezed limits. Like for the leading term, these terms agree with \cite{Flauger:2010ja} for $N=3$. The computation is completely analogous to that of the leading contribution, so we give directly the final result of leading plus first subleading term:
\be
I_{N} = (-)^{N} \sqrt{2\pi \alpha} \frac{3\alpha b H^2}{f^{N-2}} \alpha ^{N-4} \left[ \G \sin \left(\frac{\phi_K}{f} \right)- \frac{\F}{\alpha} \cos \left(\frac{\phi_K}{f} \right) \right]\,,
\ee
where $\F\equiv \sum_{i}^{N}\left(\prod _{j\neq i} k_j\right) K^{4-N}$ and we have omitted a $k$-independent phase in the argument of the trigonometric functions since this is arbitrary in the model. Putting things together we find
\be\label{fullN}
  \Expect{\prod_{i=1}^N\R_{\mathbf{k}_i}} =  (2\pi)^3\delta^3\left(\sum_{i=1}^N \mathbf{k}_i\right) \frac{A_N}{ K^{N-3} \prod_i k_i^2} \left[ \sin\left(\frac{\phi_K}{f} \right)-\frac1\alpha \cos\left(\frac{\phi_K}{f} \right) \sum_{j,i} \frac{k_i}{k_j} \right]\,.
 \ee
This is our main result.  It is natural to compare this result with the one in \cite{Flauger:2010ja} at subleading order as well. The two look very similar except that the sum in \eqref{fullN} runs over all $i$ and $j$, while in \cite{Flauger:2010ja}, e.g.~in equation (1.3), the corresponding sum runs only over $i\neq j$. The difference is given by $ - 3  \cos\left(\phi_K/f \right)/\alpha $. This term is nothing but a different choice of the $k$-independent phase omitted in the trigonometric functions in \eqref{fullN}. In order to see this, consider shifting $\phi_K/f$ in \eqref{fullN} by the $k$-independent phase $3/\alpha$. By using trigonometric relations and expanding in $\alpha\gg1$ one finds exactly the result of \cite{Flauger:2010ja}. Since the phase is anyways arbitrary, we conclude that \eqref{fullN} exactly reproduces the results of \cite{Flauger:2010ja} both at leading and at first subleading order in $\alpha$.

It is straightforward to compute all other terms in \eqref{ts}. We will use these terms in section \ref{s:aN} to estimate the combinatoric coefficients of different diagrams. The result is
\bea 
I_{N} &=& (-)^N \sqrt{2\pi \alpha} \frac{3\alpha b H^2}{f^{N-2}} \frac{\alpha ^{N-4}}{K^{N-3}}\sum_{l=0}^{N} \left(\frac{K}{\alpha}\right)^l \sin\left(\frac{\phi_{K} }{f}-l\frac{\pi}{2} \right) \sum_{i_{1}< i_{2}\dots< i_{N-l}} k_{i_{1}}k_{ i_{2}}\dots k_{ i_{N-l}}  \,,\nonumber
\eea
where $l = 0$ is the leading term. When we put this back into \eqref{ts} we find
\bea\label{allsub}
  \Expect{\prod_{i=1}^N\R_{\mathbf{k}_i}}  &=&  (2\pi)^3\delta^3\left(\sum_{i=1}^N \mathbf{k}_i\right)  \frac{A_N}{ K^{N-3} \prod_i k_i^2}\\
  &&\quad \times \left[ \sum_{l=0}^N  \sin\left(\frac{\phi_K}{f}- l\frac{\pi}{2} \right) \frac{1}{\alpha^l} \sum_{i_{1}< i_{2}\dots< i_{l}} \frac{K^l}{ k_{i_{1}}k_{ i_{2}}\dots k_{ i_{l}} } \right]\,.\nonumber
\eea
As we just discussed, the $l=0,1$ terms exactly reproduce the result of \cite{Flauger:2010ja}. On the other hand, at the next order in the large $\alpha$ expansion the calculations differ. This is presumably due to the gravitational interactions, which we have neglected and which contribute to the terms $l\geq2$. Despite this, \eqref{allsub} is still very useful to estimate the combinatorics factors arising for large $N$ as we will explain in section \ref{s:aN}.


\subsection{Squeezed Limit}

Interesting limits and consistency relations arise when some set of momenta goes to zero. 
We will be interested in two classes of consistency relations: in this subsection we consider relations arising in the \textit{squeezed limit}, i.e.~when one of the external momenta goes to zero. In the next section we will make repeated use of a relationship arising in the collinear limit, i.e.~when a (proper) subset of momenta sum up to zero. We present an extensive review of these consistency relations and how to derive them in appendix \ref{consistency}.  In the case of the squeezed limit, we can perform a non-trivial check of \eqref{fullN}. 

In the limit $k_1\ll k_2\sim k_3 \dots\sim k_N$, the $N$-point correlation function should obey
\be \label{cc}
\lim_{k_1\rightarrow 0} \Expect{\R_{\mathbf{k}_1} \cdots\R_{\mathbf{k}_N}}=(2\pi)^3\delta^3({\bf K}) |\R_{k_1}|^2\sqrt{2 \e}\frac{\partial}{\partial \phi_* }\Expect{\R_{\mathbf{k}_2} \cdots\R_{\mathbf{k}_N}}'\,,
\ee
where $\Expect{}'$ indicates that the correlator does not contain the factor $(2\pi)^3 \delta^3({\bf K})$. Although this is a relation between the full $N$- and $(N-1)$-point correlation functions, it should be satisfied order by order in an expansion in $b$ and $\alpha$. We will use the leading term in $b$ and $\alpha$ from \eqref{fullN} in both the right and left hand side of \eqref{cc}. Let us start from
\be 
\Expect{\R^N}'=A_N\frac{K^{3-N}}{\prod_i^N k_i^2}\left[\sin (a) -\frac1\alpha  \cos (a) \sum_{i,j}\frac{k_i}{k_j}\right]\,,
\ee
where we remind the reader that $a = \phi_K/f$. Then, to leading order in $k_1,b$ and $\alpha$, \eqref{cc} leads to
\be 
\frac{A_N}{A_{N-1}} \frac{1}{K k_1^2} \left[ -\frac1\alpha  \cos (a) \frac{ \sum_{i} k_i}{k_1}\right]=2 \pi^2 \frac{\Delta^2_R}{k_1^3} \sqrt{2 \e}\frac{\partial}{\partial \phi } \sin \left(\frac{\phi}{f}\right)=2 \pi^2 \frac{\Delta^2_R}{k_1^3}\alpha \cos (a)\,,
\ee
or simply 
\be 
\frac{A_N}{A_{N-1}} =- \alpha^2 \Delta_R^2 2\pi^2\,,
\ee
which is an identity given $A_N$ in \eqref{due}. Hence we have shown that the relation \eqref{cc} is satisfied to leading order in $k_1,b$ and $\alpha$ for any N.


\section{Multi-vertex Diagrams}\label{s:mv}

In the previous section we have computed all tree-level single-vertex diagrams at leading order in small $b$ using the strategy outlined in section \ref{s:gs}. The final result is given in \eqref{allsub}. Here we discuss the much harder problem of multi-vertex diagrams. We will not directly compute them, but we will estimate their size using some relations arising in the collinear limit. Details of the derivation of these relations can be found in appendix \ref{consistency}. The results collected in this section provide us with an expectation of when multi-vertex diagrams are important as compared to single-vertex diagrams. On general ground, multi-vertex diagrams are suppressed by extra factors of $b$ and $\alpha$ but their contribution can still be important, if there is a large number of terms and/or around singular points. 
We will find that for the first few correlation functions (small $N$) multi-vertex diagrams are subleading, while they grow fast for larger $N$. Combining the discussion of the current and previous sections into an accurate estimate of $N$-point correlation functions will be the subject of section \ref{s:aN}.

  \begin{figure} [ht]
\begin{center}
\includegraphics[width=0.7\textwidth, angle=0]{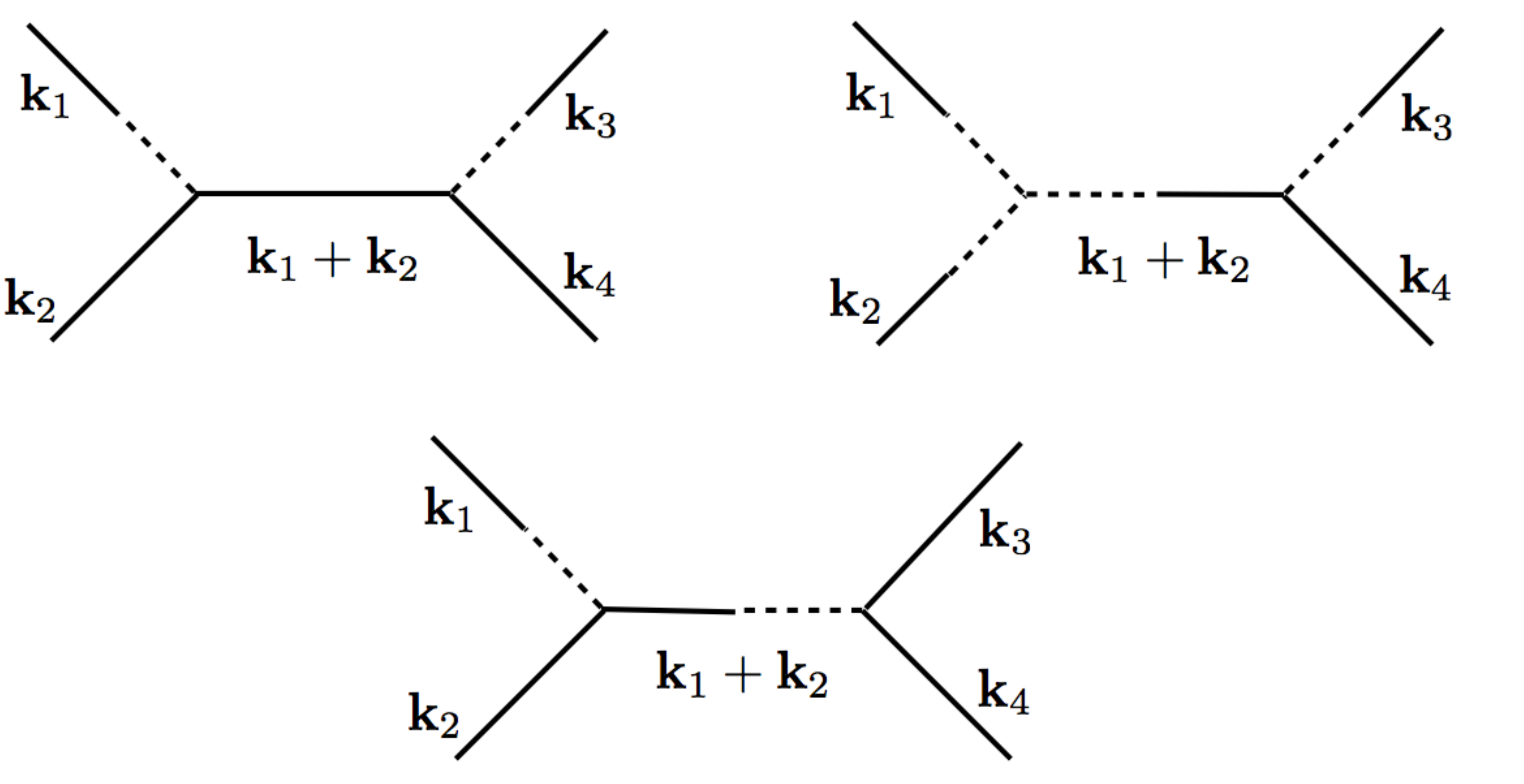}
\caption{All tree-level two-vertex diagrams for the trispectrum up to permutations. These diagrams are subleading to the one computed in figure \protect\ref{contact} for small $b$ and large $\alpha$. In the collinear limit where $\mb{q} \equiv \mb{k_1} + \mb{k_2} \rightarrow 0$ these diagrams diverge as $q^3$ and will eventually dominate over the single-vertex one in (\protect \ref{fullN}). 
}\label{split4pt}
\end{center}
\end{figure}

 \subsection{Collinear Limit with Scalar Exchange}\label{col}

The collinear limit is the limit in which a (proper) subset of the momenta $\{{\bf k}_1,{\bf k}_2,\dots,{\bf k}_l\}$ for $1<l<N$ of some $N$-point correlation function sum up to zero. In this limit, the $N$-point correlation function is dominated by diagrams describing the exchange of a long-wavelength (momentum $q\rightarrow 0$) classical scalar and tensor mode between an $l$- and an $(N-l)$-point correlation functions as depicted in figure \ref{collinear}. Here we start by analyzing the scalar exchange. In Appendix \ref{consistency} we find
\be\label{c1a}
\lim_{q\rightarrow 0}\Expect{ \prod_{i=1}^N\R_{\mathbf{k}_i} }'^{SE} = 2\e |\R_q|^2 \frac{\partial}{\partial\phi_*} \Expect{ \prod_{i=1}^l\R_{\mathbf{k}_i} }' \frac{\partial}{\partial\phi_*} \Expect{ \prod_{i=l+1}^N\R_{\mathbf{k}_i} }'\,,
\ee 
where ${\bf q}\equiv \sum_{i=1}^l {\bf k}_i $, $SE$ denotes ``scalar exchange",  and we remind the reader that the prime in $\Expect{}'$ means that the factor $(2\pi)^3\delta^3(\bf K)$ is omitted. The relation \eqref{c1a} should be verified order by order in $b$. Since we have computed the single-vertex diagrams for any $N$ we can use them in \eqref{c1a} to estimate the two-vertex $N$-point function in the collinear limit in which the propagator between the two vertices goes on shell. 

As an example let us consider $N=4$, i.e.~the trispectrum, which contains three tree-level multi-vertex diagrams depicted in figure \eqref{split4pt}. If we estimate them in the collinear limit by mean of \eqref{c1a} we find 
\be
\lim_{\mb{k_1}+\mb{k_2} \rightarrow 0} \Expect{\R_{\mb{k_1}}\cdots\R_{\mb{k_4}}}'^{SE} = 2\e |\R_{k_{12}}|^2 \frac{\partial}{\partial \phi }\Expect{\R^2_{\mb{k_1}} }'\frac{\partial}{\partial \phi }\Expect{\R^2_{\mb{k_3}} }'\,,
\ee
where $k_{ij} = |\mb{k_i} + \mb{k_j}|$. Using the power spectrum to leading order in $b$ given by \eqref{2pt}, the trispectrum in the collinear limit becomes\footnote{Notice that this expression is only valid if  $n_s -1 \ll 3b\sqrt{2\pi \alpha}$, where $n_{s}$ is the spectral index at zeroth order in $b$. This is true in the region of parameter space we are interested in (see e.g.~figure \ref{constraints}).}
\be\label{tr}
\lim_{\mb{k_1}+\mb{k_2} \rightarrow 0} \Expect{\R_{\mb{k_1}}\cdots\R_{\mb{k_4}}}'^{SE} = 9\alpha b^2 2\pi \sin^2a \frac{(2\pi^2)^3 (\Delta^2_\R)^3}{k_{12}^3k_1^3k_3^3}\,.
\ee
Two comments are in order. First, we notice that although the trispectrum depends in general on six real parameters (i.e.~$3\times4=12$ momentum components minus three for translational and three for rotational invariance), in the collinear limit, to leading order in $k_{12}$, it depends only on three, namely $k_{12},k_1,k_3$. This is because, since the exchanged scalar carries no information about angular orientation, for each of the three two-point functions separately in \eqref{tr} we can get rid of two degrees of freedom by rotational invariance (and of three by momentum conservation). So alltogether we get $6-2-3=1$ degree of freedom for each two-point function for a total of three. This will not be the case for the graviton exchange discussed in the next subsection, since the tensor propagator does carry information about the relative orientation of the vertices it connects. Second, we would like to mention that, apart for the numerical coefficients, all the scalings in \eqref{tr} could have been guessed (or computed) using the Feynman rules in appendix \ref{Feynman} (or the canonical formalism).

A similar computation can be done for higher $N$-point functions to estimate their multi-vertex diagrams close to collinear limits. For example, the five-point correlation function has one collinear divergence ($\mb{k_1}+\mb{k_2}\rightarrow 0$ plus permutations), the six-point correlation function has two collinear divergences ($\mb{k_1}+\mb{k_2}+\mb{k_3}\rightarrow 0$ and $\mb{k_1}+\mb{k_2}\rightarrow 0$ plus permutations) and so on for larger $N$'s. It is straightforward to compute the collinear limit of an $N$-point correlation function split into a subset with $N_1$ external legs and its complement. We assume that $N_1 \geq 3$ and $N_2\equiv N-N_1 \geq 3$ such that we can use \eqref{fullN}. Introducing the notation $K_N = \sum_{i=1}^N k_i$, to leading order in $b$, i.e.~to order $b^2$, the result is
\bea\label{split}
\lim_{\mb{q}\rightarrow 0} \Expect{\R_{\mb{k_1}}\cdots\R_{\mb{k_N}}   }'^{SE} & = & \alpha^2 \frac{2\pi^2 \Delta_{R}^2}{q^3}  \frac{A_{N_1}A_{N_2} }{K_{N_1}^{N_1-3} K_{N_2}^{N_2 -3} } \times \sum_{l_1}^{N_1}\sum_{l_2}^{N_2} \cos \left(\frac{\phi}{f}-l_1\frac{\pi}{2}\right)\cos \left(\frac{\phi}{f}-l_2\frac{\pi}{2}\right)\times \nonumber\\
&& \quad  \sum_{i_{1}<i_{2}\dots< i_{l_1}}  \sum_{j_{1}< j_{2}\dots< j_{l_2}}   \frac{K_{N_1}^{l_1}K_{N_2}^{l_2}} {\alpha^{l_1+l_2}\,  k_{i_{1}}k_{ i_{2}}\dots k_{ i_{l_1}}  k_{j_{1}}k_{ j_{2}}\dots k_{ j_{l_2}}} \; .
\eea
We would like to stress again that this result follows from systematically neglecting gravitational interactions. As we have seen in section \ref{ss:subl}, starting at $l=2$ some terms are missed by this approach as compared to the exact computation including gravity. Hence only the leading order in $\alpha$ terms, i.e.~$l_1=0,1$ and $l_2=0,1$ are strictly correct. On the other hand, the subleading terms in $\alpha$ in \eqref{split} should be sufficient to estimate the combinatorics factors arising for large $N$ as we will do in section \ref{s:combm}.


\subsection{Collinear Limit with Graviton Exchange}

Consistency relations together with the answer \eqref{fullN} are enough to predict the behavior of an $N$-point function in the collinear limit when the quasi on-shell internal propagator of figure \ref{collinear} corresponds to a graviton. In Appendix \ref{consistency}, we derive the consistency relation for a collinear divergence mediated by a tensor mode, the main result is
\be\label{cot}
\lim_{q\rightarrow 0}\Expect{ \prod_{i=1}^N\R_{\mathbf{k}_i} }'^{GE} = |\gamma_q|^2 \sum_{\{i,j\}=1}^{N_1 -1} \sum_{\{l,m\}=1}^{N_2-1}E_{ijlm}\frac{\partial}{\partial ({\bf k}_i\cdot {\bf k}_j)} \left\langle \prod_{i=1}^{N_1}\R_{\mathbf{k}_i} \right\rangle'  \frac{\partial}{\partial ({\bf k}_l\cdot {\bf k}_m)} \left\langle \prod_{i=1}^{N_2}\R_{\mathbf{k}_i} \right\rangle'\; ,
\ee
where $GE$ means graviton exchange and 
\be\label{cosi}
 E_{ijlm}\equiv k_ik_j k_lk_m\sin\theta_i\sin\theta_j\sin\theta_l\sin\theta_m \cos(\phi_i +\phi_j-\phi_l-\phi_m)\,,
\ee
with $\cos\theta_i \equiv \mb{k}_i\cdot \mb{q}/(qk_i)$, and $\phi_i$ is the azimuth angle on the plane perpendicular to ${\bf q}$. 
Note that the derivatives are not proportional to the spectral index and they are generally of order one. Since we have a tensor two-point function instead of a scalar one, the $GE$ contribution is suppressed\footnote{Despite this suppression, it has been argued that the collinear limit of tensor may well be the dominant piece in slow-roll model of inflation \cite{Seery:2008ax}.} with respect to the SE contribution by the tensor-to-scalar ratio $r$. Also note that in \eqref{cot} one must replace $k_N = |\sum_i^{N-1}\mb{k}_i|$ which introduces a dependence on the orientation of all momenta.
Let us see how the counting of degrees of freedom in \eqref{cot} differs from the case of scalar exchange. The left-hand side should depend on $3N-6$ real parameters (as usual due to translational and rotational invariance). On the right-hand side we have one parameter for the tensor power spectrum $q$ plus $3(N_1-1+N_2-1)$ for the angles $\theta_{i,l}, \phi_{i,l}$ and the norms $k_{i,l}$. Then we have still to subtract one parameter since adding an arbitrary constant to all $\phi_{i,l}$ leaves the result invariant due to a cancellation in the argument of the cosine in \eqref{cosi}. Altogether we have $3N-6$ parameters on the right-hand side as well. This reflects the fact that, in contrast to the scalar exchange, the graviton exchange carries information about the relative orientation of the split correlation functions and therefore, even in the collinear limit, the full dependence on the angles is preserved. 

Using \eqref{cot} for the trispectrum leads to the particularly simple result\footnote{This result was first obtained in \cite{Seery:2008ax}. There one specific direction for the collinear limit ${\bf q}\rightarrow 0$ was chosen by fixing $\theta_1=\theta_3=\pi/2$. Here we leave $\theta_1$ and $\theta_3$ unspecified. Notice that \eqref{angl} depends on six degrees of freedom, as it is the case for a general trispectrum.} (see Appendix \ref{consistency} for more details)
\bea\label{angl}
\lim_{q\rightarrow 0}\Expect{ \prod_{i=1}^4\R_{\mathbf{k}_i} }^{'GE} & =&\sin^2\theta_1\sin^2\theta_3 \cos\left[2(\phi_1 -\phi_3)\right] \frac{r}{4} \left(\frac{n_s-4}{2}\right)^2 \Expectnoc{ \R_{\mb{k}_1}^2 }'\Expectnoc{ \R_{\mb{k}_3}^2}'\Expectnoc{ \R_{\mb{k}_{12}}^2 }'\nn\\
&\simeq& \sin^2\theta_1\sin^2\theta_3 \cos\left[2(\phi_1 -\phi_3)\right] \frac{9r}{16}  \Expectnoc{ \R_{\mb{k}_1}^2 }'\Expectnoc{ \R_{\mb{k}_3}^2}'\Expectnoc{ \R_{\mb{k}_{12}}^2 }'\; ,
\eea
where we have use that $n_s\simeq1$.


\section{$N$-point Correlation Functions}\label{s:aN}

In this paper we have outlined a strategy to calculate the leading contribution to any $N$-point function in the two parameters $b\ll1$ and $\alpha\gg1$ for resonant inflationary models.  The main result is given by \eqref{fullN}. We have computed more terms \eqref{allsub} but we expect additional gravitational contributions starting at the same order as $l=2$.  
 
All corrections to this formula (higher orders of single-vertex and multi-vertex diagrams) are suppressed by extra powers of $b$ or $1/\alpha$ but might be enhanced by two effects. First, in some special part of k-space, a large ratio of momenta might arise, such as in collinear limits. This is not a serious obstacle because exactly in these regimes we can make use of consistency relations. Second, the number of subleading diagrams grows with $N$ making their sum a potentially large contribution even if each one of them individually is very small. This is related to a well known problem in particle physics, where calculating scattering amplitude for large number of external legs is made computationally challenging by the fast growing number of Feynman diagrams. New techniques are being devised to handle such calculations, e.g.~in gauge theories, and it might be interesting to investigate whether any one of them could be employed in the context of cosmological perturbations. 

In the following, using the results of the previous two sections, we estimate both of these effects. We find that combinatoric factors grow fast and can overcome the $b$ and $\alpha$ expansion for sufficiently large $N$. We estimate the largest $N$, which we call $\NN$, for which the $b$ and $\alpha$ expansion are trustable (summarized in table \ref{t1}). For $N<\NN$, all correlation functions can be accurately computed using the results given in this paper. In the bulk of momentum space, i.e.~for generic momenta, the result is given by \eqref{fullN}. In certain localized regions of momentum space, which we indicate as collinear regimes, consistency relations can be used to compute the leading behavior. As an example we do this in detail for the trispectrum in subsection \ref{ss:ts}. 

Before proceeding, let us stress that the fact that we are not able to compute $N$-point correlation functions above a certain $\NN$ does not imply a failure of the general strategy outlined in section \ref{s:gs}. The problem lies somewhere else. Even in the relatively simple scalar theory obtained after neglecting gravitational interactions, very large $N$-point correlation functions are computationally intractable.


\subsection{Estimate of the combinatorics of single-vertex diagrams}

As we have seen in subsection \ref{ss:subl}, only the leading and first subleading order in $\alpha$ of our computation of single-vertex diagrams \eqref{fullN} correctly reproduces the exact computation including gravity. Hence our result is trustable as long as the higher orders are indeed negligible. Although these terms are suppressed by powers of $\alpha$ they are enhanced by combinatoric factors which grow with $N$. In the following we estimate the smallest $N$, which we will call $\NNsv$, for which the $l=2$ subleading terms start to dominate over the $l=0,1$ terms of \eqref{allsub}.

Using \eqref{allsub}, we take the ratio of the subleading terms in $\alpha$, i.e.~those with $l>2$, to the leading term plus first subleading term in $\alpha$, i.e.~the one with $l=0$ and $l=1$. To simplify the estimate, we will assume that all $k$'s are of the same order, so that $K=\sum_i^N k_i\simeq N k$. Once this is done, one finds
\bea 
| l=0,1 \textrm{ leading single-vertex}|&\propto & 1- \frac{N^2}{\alpha} \cos\left(\frac{\phi_K}{f}\right)= \nonumber  \sqrt{1+\frac{N^4}{\alpha^2}} \sin\left(\frac{\phi_K}{f}+\mathrm{phase}\right)\,.
\eea
A similar simplification can be done for the sum of $l>2$ terms to write them all as a single overall amplitude times a sine up to a phase. Since this leads to cumbersome formulae, we prefer to simply assume that all $l>2$ terms interfere constructively, i.e.~we approximate $\sin\sim\cos\sim 1$ in summing them up. This gives a conservative estimate of $\NN$, which we have checked is practically undistinguishable from the one obtained by correctly accounting for all trigonometric functions. So our estimate goes as follows
\bea\label{r1}
\frac{|\rm l >2 \textrm{ subleading single-vertex}|}{| l=0,1 \textrm{ leading single-vertex}|} &=& \frac{ \sum_{l=2}^N \frac{K^l}{\alpha^l} \sum_{i_{1}< i_{2}\dots< i_{l}}( k_{i_{1}}k_{ i_{2}}\dots k_{ i_{l}})^{-1} \sin\left(\frac{\phi_K}{f}-l\frac{\pi}{2}\right)}{1- \frac1\alpha \sum_i \frac{K}{k_i}\cos(\frac{\phi_K}{f})}\nonumber \\
&\sim&\frac{\sum_{l=2}^N\left(\frac{N}{\alpha}\right)^l\binom{N}{l}}{\sqrt{1+ N^4/\alpha^2}}=\frac{\left(1+\frac{N^2}{\alpha}\right)^N-1-\frac{N^2}{\alpha}}{\sqrt{1+ N^4/\alpha^2}}\,,
\eea
where the binomial coefficient is given by $\binom Nl\equiv N!/[l!(N-l)!]$. The first $N$ and $l$ for which this quantity becomes of order one depends on $\alpha$. 
Both theoretical and phenomenological arguments guide our attention to values of $\alpha$ in the range $100-1000$. Let us go through these arguments.

For an inflationary background of the large-field type as in \cite{McAllister:2008hb,Flauger:2009ab,Berg:2009tg}, we expect $\sqrt{2\e}\sim\mathcal{O}(.1)$, while for $f$ (remember in our conventions we have set the reduced Planck constant to one) we expect some subPlanckian value, i.e.~$f\ll1$ \cite{Banks:2003sx}. Since $b$ is related to some non-perturbatively generated scale $\Lambda$, we assume $b\lesssim.1$. On the phenomenological side, we find again that $b\ll1$ is required by comparison of the resonant spectrum with the data \cite{Pahud:2008ae,Flauger:2009ab}. Then, the resonant bispectrum can be estimated by \cite{Chen:2008wn,Hannestad:2009yx,Flauger:2010ja} $f_{\rm res}=3b\sqrt{2\pi}\alpha^{3/2}/8\simeq b \alpha^{3/2}$. Therefore a detectable bispectra will require $\alpha$ of order $10^2$ or larger. Putting these observations together we conclude that an interesting range of values is $\alpha\sim 10^2-10^3$. Demanding that \eqref{r1} is less than $20\%$ gives $\NNsv$ for a given $\alpha$. Some explicit values are
\be 
\{\alpha=100,\NNsv=9\},\{\alpha=500,\NNsv=18\},\{\alpha=1000,\NNsv=26\}\,.
\ee


\subsection{Estimate of the combinatorics of multi-vertex diagrams}\label{s:combm}

The leading order contribution \eqref{fullN} only includes the single vertex diagrams depicted in figure \ref{contact}. Each vertex comes with a factor of $b$ and therefore the amplitude of multi-vertex diagrams is subleading in that parameter. Also, although it is not as obvious, the amplitude is subleading in $\alpha$. In the following we estimate the combinatoric factors associated to these diagrams which could invalidate the $b$ and $\alpha$ expansions. We do not directly compute any multi-vertex diagram, but we use the results of section \ref{s:mv} for the collinear limit to estimate them. Our final result will be an estimate of $\NNmv(\alpha,b,r)$, i.e.~the largest $N$ for which the $b$ and $\alpha$ expansions are not invalidated by the large number of multi-vertex diagrams. 

We start by considering only the case of scalar exchange (SE)
and we look at the multi-vertex diagrams arising from splitting the $N$-point correlation function into $N_1$ external momenta on the one side and $N_2\equiv N-N_1$ on the other. For simplicity we take all momenta to be the same $k_i \simeq k$ and adopt the notation $K_N = \sum_{i=1}^N k_i = N k$. We also assume that $N_1 \geq 3$ and $N_2 \geq 3$ such that we can use \eqref{fullN}. One can check that the case $N_1=2$ and $N_{2}=2$ give a smaller contribution.  We then find
\bea
\frac{\lim_{\mb{q}\rightarrow 0} \Expect{\R_{\mb{k}_1}\cdots\R_{\mb{k}_N} } ^{'SE}}{\Expect{\R_{\mb{k}_1}\cdots\R_{\mb{k}_N} }'} & = & \frac{2\e |\R_q|^2 \frac{\partial}{\partial \phi_* }\Expect{\R_{\mb{k}_1} \cdots \R_{\mb{k}_{N_1}} }'\frac{\partial}{\partial \phi_* }\Expect{\R_{\mb{k}_{N_1+1} }\cdots\R_{\mb{k}_N} }' } {| l=0,1 \textrm{ leading single-vertex}|}\nonumber\\
& \sim & \alpha^2 (2\pi^2 \Delta_{R}^2) \frac{A_{N_1}A_{N_2}}{A_{N_1+N_2}} \left(1+\frac{N^4}{\alpha^2}\right)^{-1/2} \frac{K_N^{N-3}}{K_{N_1}^{N_1-3} K_{N_2}^{N_2 -3} q^3}\nonumber\\
& = & \frac{3b\sqrt{2\pi}}{2\alpha^{5/2}} \frac{N^{N-3}}{N_1^{N_1-3} N_2^{N_2-3}} \left(1+\frac{N^4}{\alpha^2}\right)^{-1/2}\frac{k^3}{q^3}\,.\label{co2}
\eea
The numerical coefficient grows quickly with $N$ and is maximized for $N_2=N_1= N/2$ where it is $N^3 2^{N-6}\left(1+N^4/\alpha^{2}\right)^{-1/2}$. This is expected since our leading contribution is a single term in k-space while as many as $N^3$ terms contributes to the collinear limits.

An analogous estimate can also be obtained for the collinear multi-vertex diagrams with graviton exchange (GE).  Starting with \eqref{cot}, let us focus for concreteness on the $N_1$ part, since the $N_2$ will be identical. Again we assume $N_1,N_2\geq3$, since we checked that $N_1=2$ and $N_{2}=2$ give a smaller contribution. We notice that the derivatives are with respect to ${\bf k}_i\cdot {\bf k}_j$ with $i,j=1,\dots,(N_{1}-1)$, meaning that ${\bf k}_{N_{1}}$ should be re-expressed in terms of the other $(N_{1}-1)$ momenta as  $ {\bf k}_{N_{1}} =-\sum_{i}^{N_{1}} {\bf k}_i$. The dominant term in \eqref{cot} arises when the derivative acts on $\sin(a)$ inside $\Expect{\R^{N_{1}}}$ giving 
\be
\frac{\partial} {\partial ({\bf k}_i\cdot {\bf k}_j)}\sin(a)=\frac{\partial}{\partial ({\bf k}_i\cdot {\bf k}_j)}\sin\left(\frac{\phi_*}{f}-\alpha \log \frac{K}{k_*}\right)=-\frac{\alpha}{2K}\left[\frac{\delta_{ij}}{k_i}+\frac{1}{k_N}\right]\cos(a)\,,
\ee
for any $i,j=1,\dots,(N_{1,2}-1)$. Contracting this with $E_{ijkl}$ introduces a complicated angular dependence. We decide to give a very conservative estimate, by neglecting the cancellations occurring due to the various sinusoidal dependences and setting $\sin \sim \cos \sim \mathcal{O}(1)$. This gives
\bea
\sum_{i,j}^{N_1-1} {\bf k}_i^T\epsilon^s{\bf k}_j\frac{\partial}{\partial ({\bf k}_i\cdot {\bf k}_j)} \Expect{\prod_{i=1}^{N_1}\R_{\mathbf{k}_i}}' &\sim& -\sum_{i,j}^{N_1-1}   k_ik_j \frac{\alpha}{2K}\left[\frac{\delta_{ij}}{k_i}+\frac{1}{k_N}\right]\cos(a)  \Expect{\prod_{i=1}^{N_1}\R_{\mathbf{k}_i}}' \nonumber\\
&\sim&- \frac{\alpha}{2} (N_1-1)  \Expect{\prod_{i=1}^{N_1}\R_{\mathbf{k}_i}}'\,,
 \eea
where in the last step we have assumed for simplicity that all momenta are of the same order, $k_i\simeq k$. Putting things together our estimate for \eqref{cot} is
\be
\lim_{q\rightarrow 0}\Expect{ \prod_{i=1}^N\R_{\mathbf{k}_i} }'^{GE} \sim 2  \frac{r}{4}\frac{\alpha^2}{4} |\R_q|^2 (N_1-1) (N_2-1) \Expect{\prod_{i=1}^{N_1}\R_{\mathbf{k}_i}}'  \Expect{\prod_{i=N_{1}+1}^{N}\R_{\mathbf{k}_i}}'\,,
\ee
where we have introduced the tensor-to-scalar ratio defined in \eqref{r} and the extra factor of two comes from the sum over polarizations. We therefore obtain the result
\bea
\frac{\lim_{\mb{q}\rightarrow 0} \Expect{\R_{\mb{k}_1}\cdots\R_{\mb{k}_N}}^{GE} }{\Expect{\R_{\mb{k}_1}\cdots\R_{\mb{k}_N}}} &\sim & \frac{\alpha^2 r |\R_q|^2  (N_1-1) (N_2-1)\Expect{\R_{\mb{k}_1} \cdots \R_{\mb{k}_{N_1}}}'\Expect{\R_{\mb{k}_{N_1+1}}\cdots\R_{\mb{k}_N} }'}{8 | l=0,1 \textrm{ leading single-vertex}|}\nonumber\\
& \sim& \frac{\alpha^2 r}{8}\frac{(N_1-1) (N_2-1) (2\pi^2 \Delta_{R}^2)}{ \sqrt{1+\frac{N^4}{\alpha^2}}} \frac{A_{N_1}A_{N_2}}{A_{N_1+N_2}}\frac{K_N^{N-3}}{K_{N_1}^{N_1-3} K_{N_2}^{N_2 -3} q^3}\nonumber\\
& \sim & \frac{3b\sqrt{2\pi}}{16\alpha^{5/2}} \frac{r(N_1-1) (N_2-1)}{\sqrt{1+\frac{N^4}{\alpha^2}}} \frac{N^{N-3}}{N_1^{N_1-3} N_2^{N_2-3}} \frac{k^3}{q^3}\label{co3}\,.
\eea
The numerical coefficient is maximized for $N_1 = N_2 = N/2$ where the combinatorics is $(N/2-1)^2 N^3 2^{N-6}\left(1+N^4/\alpha^{2}\right)^{-1/2}$. Comparing with \eqref{co2}, we see that the combinatorics is worse than for the scalar exchange by a factor $(N/2-1)^2$ but there is an extra suppression by a factor of $r$ which is at most a fifth but could be much smaller. We remind the reader that we have assumed total constructive interference between various oscillating terms and therefore this is a very conservative overestimate. 


\subsection{$N$-point Correlation Functions}

Now that we have either computed or estimated all types of contributions to a generic tree level $N$-point function, we are ready to collect our findings into a final result. We will first discuss a generic $N$-point function away from collinear limits, i.e.~away from the localized regions in momentum space where a combination of momenta vanishes\footnote{Our formula also does not capture multiple squeezed or multiple collinear limits where more than one external momentum (or more than one internal momentum) go to zero. 
We have not attempted to study this case thoroughly and we leave this step for future investigation. It should be noted that multi-squeezed/collinear-divergent terms become large only in very small regions of momentum space, which have effectively codimension two or more. 
}. 
Thanks to consistency relations we will be able to return to those collinear limits later on. 

\begin{table}[ht]
\centering 
\begin{tabular}{|c|c||c|c|c||c|c|c|}
\hline
\multicolumn{2}{|c||}{\textbf{Single vertex}} & \multicolumn{3}{|c||}{\textbf{Multi-vertex Scalar}} & \multicolumn{3}{|c|}{\textbf{Multi-Vertex-tensor}} \\
\hline
$\alpha \backslash b$ & any & $\alpha\backslash b$ & 0.1 & 0.01 &$\alpha\backslash b$ & 0.1 & 0.01 \\
\hline
 100 &  9 & 100 & 12 & 15 & 100& 13 & 16\\
\hline 
1000 & 26 & 1000 & 18 & 21 & 1000 & 18 & 21  \\
\hline
\end{tabular}\label{t1}
\caption{Estimate of the value $\NN$ for different cases. For single vertex, $\NN$ is defined as the smallest $N$ for which the $b$- and $\alpha$-subleading terms ($l>2$ in (\protect \ref{allsub})) add up to $20\%$ or more of the $b$- and $\alpha$-leading terms given in (\protect \ref{fullN}). For the collinear scalar and graviton exchange, $\NN$ is defined as the smallest $N$ for which the conservative estimate of the combinatorics in (\protect \ref{co2}) and (\protect \ref{co3}) with a ratio of scales $k/q\sim1$ add up to  $20\%$ or more of the $b$- and $\alpha$-leading terms.  For the graviton-exchange diagrams we have taken the tensor to scalar ratio $r$ to be 0.1 in order to give a conservative estimate. For inflationary backgrounds of the small-field type $r$ could be much smaller.}
\label{t1}
\end{table}

Away from collinear divergences, three quantities control the size of a given diagram: $b$, $\alpha$ and the combinatorics\footnote{In principle one should also include the tensor-to-scalar ratio $r$ appearing in the contribution from graviton exchange. We provide a conservative estimate assuming $r\sim.1$, which is the largest order of magnitude allowed by the data. Our results are modified very little if one assumes a different fiducial value for $r$.}. In section \ref{s:sv} we have obtained the leading order result in $b$ and $\alpha$, given in \eqref{fullN} (with coefficient given by \eqref{tre}).  Although this result was obtained neglecting gravity, it is a very accurate approximation to the exact result because gravitational interactions would give small corrections at the same order in $b$ and $\alpha$. Intuitively this is because the relevant interactions come from the scalar field sector, while the gravitational ones are much smaller. Therefore, as long as $b\ll1$ and $\alpha\gg1$ are good expansion parameters any $N$-point function away from collinear limits is given by \eqref{fullN}.

\begin{figure} [ht]
\begin{center}
\includegraphics[height=.6\textwidth]{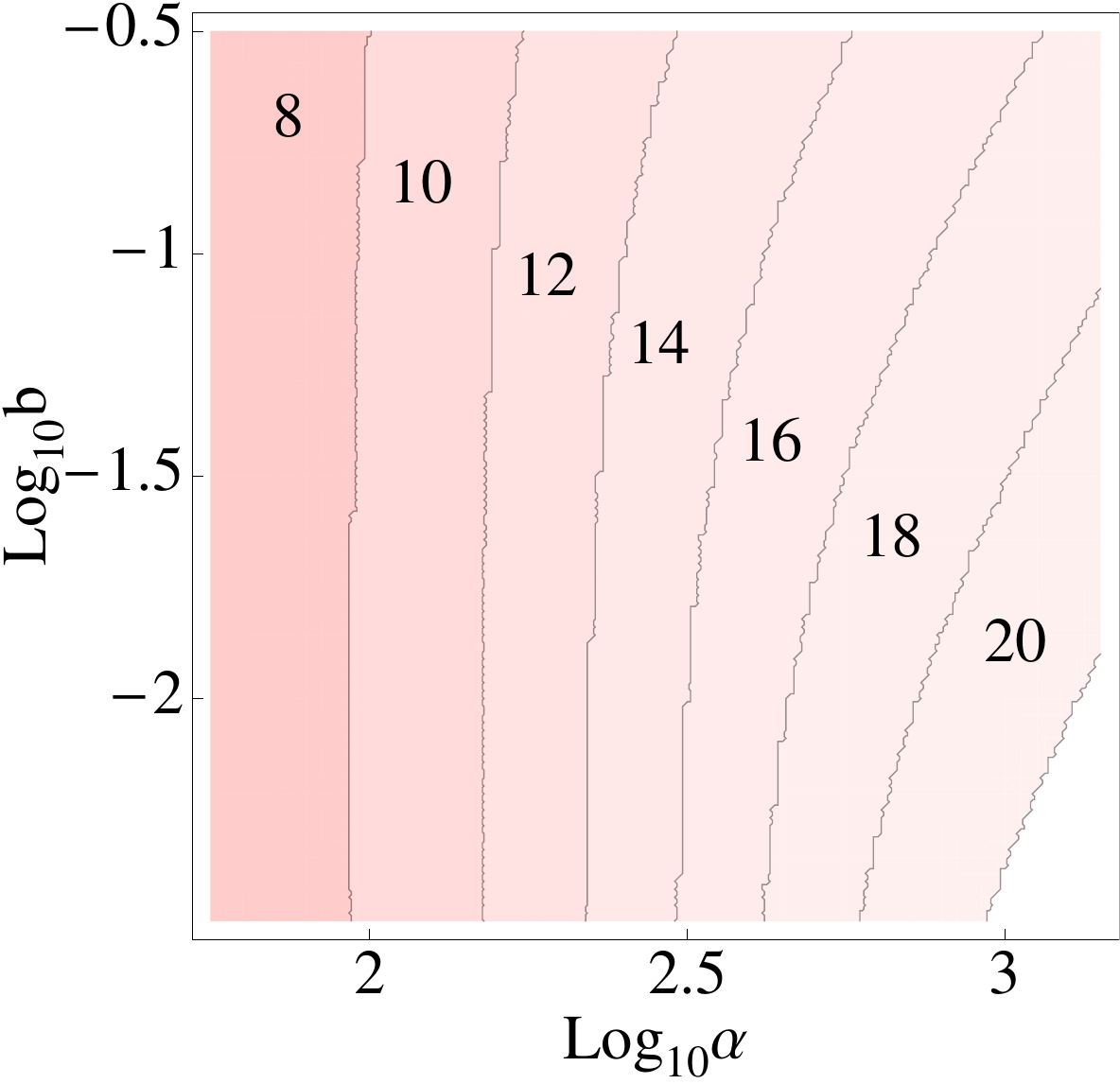}
\caption{The plot shows $\NN$ for different regions of the $\log_{10}(b)-\log_{10}\alpha$ plane. As expected smaller $b$ and larger $\alpha$ extend the validity of (\protect \ref{fullN})-(\protect \ref{tre}) to higher and higher $N$'s.}\label{constraints}
\end{center}
\end{figure}

Due to the fast growth of combinatoric factors for $N>\NN$, $b$ and $\alpha$ cease to be good expansion parameters, i.e.~higher orders are not negligible anymore. In the two previous subsections, we have estimated this growth for some $b$- and $\alpha$-subleading diagrams. The smallest $N$ for which the combinatoric factors overcome the $b$ and $\alpha$ suppression are collected in table \ref{t1} for various diagrams and values of the parameters. For concreteness we have written down the smallest $N$ for which the $b$- and $\alpha$-subleading terms from each class of diagrams are $20\%$ or more of the $b$- and $\alpha$-leading terms\footnote{Note that by leading terms we mean all the terms in \eqref{fullN}. It includes the first $1/\alpha$ corrections.}. The smallest $N$ for which the sum of \textit{all} subleading terms is $20\%$ or more of the $b$- and $\alpha$-leading terms is $\NN$. This is visualized in figure \ref{constraints}. For example for $\alpha=100$ and any $b<.1$, $\NN=9$, while for $\alpha=1000$ and any $b<.1$, $\NN=17$. To summarize, \textit{\eqref{fullN} is a very accurate approximation to all $N$-point correlation functions for $2<N<\NN$ away from collinear limits.}

Let us now turn to the discussion of collinear limits. The momentum dependence of any $N$-point function is such that divergences arise when some momenta or a combination of them go to zero, familiar examples being the $k^{-3}$ scaling of the spectrum of a free field in de Sitter or of local non-Gaussianity in the squeezed limit. In our computation, one might be worried that when such divergences appear in a subleading term, the latter can become very large and invalidate our result. Luckily for us, the squeezed limit in which a {\it single} external momentum goes to zero is correctly captured by the first subleading term in \eqref{fullN}. On the other hand, collinear and multiple squeezed enhancement appears in subleading terms that our method cannot compute. Yet again, we are lucky since anytime
these singularities are present, the $N$-point function can be rewritten in terms of lower-$N$-point functions away from these special point using the consistency relations reviewed in appendix \ref{consistency}. Therefore it is straightforward to include these collinear singularities into our result  \eqref{fullN}. As an example, we include collinear singularities for the trispectrum in the next section.


\subsection{The Trispectrum}\label{ss:ts}
Even if the combinatorics are under control, one expects that the leading formula \eqref{fullN} will fail near collinear singularities. It would be interesting to have a better approximation of the $N$-point correlation functions that is valid everywhere in k-space. This is in principle easy to do since we can use the consistency relations to find out the correct behavior, with coefficients, in each of these collinear limits. In practice, it is very cumbersome to write down all the possible terms at large N but in a given application, one will usually need only a specific limit which is easy to figure out from our formula in case by case situations. Here we will give the trispectrum including the corrections in the collinear limits (both from scalar and tensor exchange). 

A very good approximation to the trispectrum can be obtained by taking our leading results \eqref{fullN} (with $ a=\phi_K/f $)  
\bea
 \Expect{ \R_{\mathbf{k}_1}\cdots \R_{\mathbf{k}_4} }  &=& (2\pi)^3 \delta^3\left(\sum_{i=1}^4 \mathbf{k}_i\right) \frac{A_4}{ K \prod_i k_i^2} \left[ \sin\left(a\right)-\frac1\alpha \cos\left(a\right) \sum_{j,i} \frac{k_i}{k_j} \right]\label{tribasic}\,,\\
 A_4 &=& \frac32\sqrt{2\pi} b \alpha^{7/2} (2\pi^2\Delta^2_R)^3\,,\nonumber
\eea
and adding the relevant momentum singularities using the various consistency relations. For the four point function, we have collinear divergences mediated by scalars and tensors as well as squeezed divergences. The latter is already included in \eqref{tribasic}. We also have multi-squeezed-limits and the case of collinear and squeezed limits taken at the same time. We will not include them in the present discussion as the necessary consistency relations are harder to derive. We leave the analysis of these terms for further work, noticing that they are relevant only in very localized regions and do not affect the bulk shape of the $N$-point function. Using the consistency relation in appendix \ref{consistency}, \eqref{fullN} for the $N\geq3$ and \eqref{2pt} for $N=2$ we find
\bea
 \Expect{ \R_{\mathbf{k}_1}\cdots \R_{\mathbf{k}_4} }^{\rm{collinear}} & =  &(2\pi)^3 \delta^3\left(\sum_{i=1}^4 \mathbf{k}_i\right) \left\{\frac{A_4}{ K \prod_i k_i^2} \left[ \sin\left(a \right)-\frac1\alpha \cos\left(a\right) \sum_{j,i} \frac{k_i}{k_j} \right]\right.+\nn\\
 &&\left.+ \frac{9}{16}r \sin^2\theta_1\sin^2\theta_3\cos\left[2(\phi_1+\phi_3) \right] \frac{ (\td)^3}{k_{12}^3k_1^3k_3^3} \right.\nn\\ &&\left.+ \frac{18 \pi b^2\a\sin^2(a)(\td)^3}{k_{12}^3 k_1^3 k_3^3} 
 + \textrm{perm} (k_{13}, k_{23}) \right\}\; .\label{tri}
\eea
Note again that we have not included the double squeezed limit and that this formula is \emph{only} valid near collinear singularities, i.e.~when any one of $k_{1,2},k_{1,4},k_{2,3}$ goes to zero. One should not take a squeezed limit of this expression, the answer in the squeezed limit is given by \eqref{tribasic}. The size of the various terms can be estimated by assuming that all momenta are of the same size except the vanishing ones. One finds
\begin{align}
\frac{\rm{squeezed}}{\alpha\; \rm{leading}} & = -\frac{4}{\alpha}\frac{\cos(a)}{\sin(a)} \frac{k}{k_j} \; ,\\
\frac{\rm{collinear-scalar}}{\alpha\; \rm{leading}} & = \frac{24\sqrt{2\pi} b \sin(a)}{\alpha^{5/2}} \frac{k^3}{k_{12}^3}\; , \\
\frac{\rm{collinear-tensor}}{\alpha\; \rm{leading}} & = \frac{3 r \sin^2\theta_1\sin^2\theta_3\cos(2(\phi_1+\phi_3) )}{2\sqrt{2\pi} b \alpha^{7/2}\sin(a)} \frac{k^3}{k_{12}^3} \; \label{tricol}
\end{align}
where $k_j$ and $k_{12}$ are going to zero. We see that all these terms are very small in the equilateral limit, since the combinatorics are not yet important here, but they are enhanced by a large ratio of scales near collinear singularities. For CMB observations this ratio can be as large as $k/k_j \sim k/k_{12} \sim 10^3$.
We can estimate the size of the trispectrum by evaluating it at different points, say the equilateral, collinear and squeezed points,
\bea
t_{NL} &\equiv&  \frac{\Expect{ \R_{\mathbf{k}_1}\cdots \R_{\mathbf{k}_4} }'|_{\rm{equilateral}}}{|R_k|^6}\; ,\\
\tau_{NL}& \equiv &\frac{\Expect{ \R_{\mathbf{k}_1}\cdots \R_{\mathbf{k}_4} }'|_{k_{12} \rightarrow 0}}{|R_{k_{12}}|^2|R_{k_{1}}|^2|R_{k_{3}}|^{2}}\; ,\\
g_{NL} & \equiv & \frac{25}{54}\frac{\Expect{ \R_{\mathbf{k}_1}\cdots \R_{\mathbf{k}_4} }'|_{k_{1} \rightarrow 0}}{|R_{k_{1}}|^2|R_k|^4}
\eea
where $|R_k|^2$ is the power spectrum define in Eq.~\eqref{2pt}. Note that our definition of $t_{NL}$ differs from the one in \cite{Chen:2009bc} by a factor of 8 while $\tau_{NL}$ and $g_{NL}$ agrees with the literature  \cite{Byrnes:2006vq} when taking the appropriate limits. Evaluating our trispectrum in the appropriate limits using Eq.~\eqref{tribasic} almost everywhere and Eq.~\eqref{tri} in the collinear limits we get
\bea
t_{NL} &= &\frac{3\sqrt{2\pi} b\alpha^{7/2}\sin a}{8}\; ,\\
\tau_{NL}&=& \frac{9}{16}r\sin^2\theta_1\sin^2\theta_3\cos\left[2(\phi_1+\phi_3) \right] +18 \pi b^2\alpha \sin^2(a)\; ,\\
 g_{NL} &= &-\frac{75}{108}\sqrt{2\pi} b \alpha^{5/2} \cos a\; .
\eea
Note that we have not specified exactly how one takes the limit $k_{12} \rightarrow 0$ to get $\tau_{NL}$ which is why there is still an angular dependence. Current bounds on the trispectrum can be found in \cite{Vielva:2009jz}.
Since our signal is oscillatory, probably a dedicated analysis is requires to extract it from the data \cite{list, Pahud:2008ae}. Notice that in writing the above formulae we are omitting an arbitrary phase in $\sin (a)$ and $\cos(a)$, which is not fixed by the model.


\section{Discussion}\label{s:dc}

In this paper we have computed for the first time the $N$-point primordial curvature correlation functions from inflation up to $N$ of order ten or more at tree level. We did this in resonant inflationary models \cite{Chen:2008wn} with the potential \eqref{V}, which find a natural UV completion in models based on axion monodromy \cite{Silverstein:2008sg} constructed in string theory \cite{McAllister:2008hb,Flauger:2009ab,Berg:2009tg}. Our main results is
\bea \label{last}
  \Expect{\R_{\mathbf{k}_1} \cdots\R_{\mathbf{k}_N}} \Big|_{N<\NN}&=&  (2\pi)^3 \delta^3\left(\sum_i^N{\bf k_i}\right) A_N \B \,, \\
 A_N &\equiv& (-)^N \frac{3 b \sqrt{2\pi}}{2} \alpha ^{2N-9/2}\Delta_R^{2N-2}(2\pi^2)^{N-1}\nonumber \,,\\
   \B(k_i)& \equiv &\frac{1}{ K^{N-3} \prod_i k_i^2} \left[ \sin\left(\frac{\phi_K}{f} \right)-\frac1\alpha \cos\left(\frac{\phi_K}{f} \right) \sum_{j,i} \frac{k_i}{k_j} + \mathcal{O}(\alpha^{-2})\right]\; .\nn
 \eea
This formula captures the leading behavior of the N-point correlation away from collinear singularities for $N<\NN$, where $\NN$ is of order ten or larger depending on the value of the parameters. Corrections to this formula are less than 20\% for any $N<\NN$ (see figure \ref{constraints} for the list of $\NN$). Let us review the most important ingredients of our calculation. 

First, we have noticed that the main interactions in our inflationary model come from the scalar field sector and can in principle be made parametrically large (by taking a large $\alpha$). This justifies neglecting gravity, which drastically simplifies the computation. We have checked that our result is an excellent approximation to the exact result including gravity, reproducing it for $N=3$ at leading and first subleading order in $\alpha$. The success of this decoupling strategy was already noticed several times in the literature and ours is an additional motivation to formalize it for a generic case. 

Even after the problem has been simplified to computing correlation functions for a relatively simple scalar field theory, some further obstacles are encountered. The first is the presence of collinear limits (or multiple squeezed/collinear limits). The issue is that even very small contributions to a correlation function, which are potentially sensitive to gravitational interactions, could in principle be boosted by momentum enhancement arising in some particular limits. Since our strategy is to decouple and consequently neglect gravity, these terms seem hard capture. Fortunately, consistency relations exist that relate squeezed/collinear limits to correlation functions with less external legs, the latter evaluated away from these points. The second obstacle is the fast growth with $N$ of combinatoric coefficients. This is a well known problem in field theory and in fact a very active field of research. We have not been able to overcome this obstacle which therefore sets the limits of validity of our computation defining an $\NN$ above which our result \eqref{last} is not trustable. 

Since our approach has been partially inspired by the effective field theory of perturbations around an inflationary background proposed in \cite{Cheung} (and recently generalized to multiple fields in \cite{Senatore:2010wk}) let us comment on the connection. The authors of \cite{Cheung} start from the most general Lagrangian for perturbations around a homogenous time dependent inflationary background. In order to recover time diffeomorphism invariance one can perform the St\"uckelberg trick introducing a new field $\pi$, which non-linearly realizes the broken symmetry. This formulation stresses that just the effective interactions amongst the perturbations are relevant for actual observables so that one is able to differ a precise construction of what generates the inflationary background. When the underlying theory is specified as in the resonant model considered in this paper, then it is straightforward to switch from the standard description of perturbations to the $\pi$ gauge. Independently of one's favorite variables, the important point is that there is a decoupling limit in which the metric can be approximated as unperturbed. For observables not dominated by gravitational interactions this limit can be used (as we did) in order to drastically simplify computations.  To clarify the equivalence between the description of \cite{Cheung} and the one used in the present paper, let us briefly sketch how our computation translates into the $\pi$ variable. Since our model has a canonically kinetic term and no higher derivatives\footnote{The simultaneous presence of modulations in the potential and higher derivatives has been studied in \cite{Chen:2010bka} using the $P(X,\phi)$ formalism.} we can set $M_2 = M_3 = 0$ in \cite{Cheung}. One can then follow the derivation in \cite{Cheung:2007sv} and solve the constraints of the most general gauge-unfixed Lagrangian for $g_{\mu\nu}$ and $\pi$ using the ADM formalism. At linear order one finds\footnote{This result was derived in \cite{Maldacena:2002vr} both in comoving and spatially flat gauge. In the presence of higher powers of single derivatives the right hand side of \eqref{mocons} is simply multiplied by $c_{s}^{-2}$, while the constraint for $N$ is unchanged.} 
\bea
\delta N&=&\epsilon H \pi\,, \\
\partial_{i}N^{i}&=&-\epsilon H \dot \pi+\epsilon^{2} H^{2} \pi\,. \label{mocons}
\eea
No slow-roll approximation was used to obtain this result. In the action there are many couplings between $N$, $N^i$, the metric perturbations $\gamma_{ij}$ and $\pi$. Substituting the value of the lapse and shift found above into the action, one can systematically check that the kinetic term $(\partial\pi)^2$ dominates over everything else at the quadratic level. This is because in resonant models $\epsilon\ll1$, so that, at least at the level of linear constraints, decoupling works as in standard slow-roll inflation i.e. one can simply study the decoupled $\pi$ Lagrangian approximating the metric as unperturbed. Then the crucial point is to keep higher derivatives of the time dependent coefficients (in our case only $H$ and its derivatives) when expanding in small $\pi$. For example for the bispectrum the relevant terms is the one with the largest number of derivatives, i.e. $-H^{(4)}\pi^{3}/6$. To leading order in $\alpha$ this is the same as $V'''\delta \phi^3/6$ given the relation $\delta \phi=-\pi \dot\phi$ valid at linear order.

A systematic term by term inspection of the decoupling limit can also be done using the $\delta\phi$ action derived in \cite{Maldacena:2002vr}. Every terms in Eq.~(3.8) of \cite{Maldacena:2002vr} except the $V'''$ term are gravitational in origin. By inspection, they are all subleading with respect to $V'''$ for resonant inflationary models.

To conclude, we mention some interesting directions for future research:
\begin{itemize}
\item Determine a general formulation of the decoupling limit and provide some formal proof of its validity.
\item So far all examples seem to confirm the validity of the strategy outlined in section \ref{s:gs}. Determine whether there are counterexamples. Can deviations from Bunch-Davies  \cite{Chen:2006nt, Meerburg:2009ys, Chen:2010bka} be seen as single field models with parametrically large non-Gaussianity? Can we have a simple quantitive test of whether a theory decouples or not? Models with features in the potential are also single field models with parametrically large non-Gaussianity, can the decoupling limit be applied to them? 
\item Investigate in detail consistency relations for multiple independent squeezed/collinear limits.  
\item Use our results to perform a detailed phenomenological analysis of the trispectrum (and higher point functions) for resonant models.
\item Investigate the mechanism of resonant enhancement for isocurvature perturbations.
\end{itemize}


\section*{Acknowledgments}

We are grateful to Niayesh Afshordi, Brando Bellazzini, Latham Boyle, Xingang Chen, Paolo Creminelli, Raphael Flauger, David Seery, Sarah Shandera, Filippo Vernizzi and Scott Watson for useful discussion. L.L is grateful to the Centro de Ciencas de Benasque ``Pedro Pascual" and the Center for Cosmology at DAMTP for their hospitality while this work was completed. Research at the Perimeter Institute is supported in part by the Government of Canada through NSERC and by the Province of Ontario through the Ministry of Research and Information (MRI).The research of E.P. was supported in part by the National Science Foundation through grant NSF-PHY-0757868.


 \appendix 


 \section{Feynman rules}\label{Feynman}
 
 In this appendix, following \cite{vMS} we derive the Feynman rules for the theory for the perturbations that we study in the main text. The expectation value of an operator Q at time $t > t_i$ that starts in the $\Ket{in}$ state subject to a time dependent Hamiltonian $H(t)$ is 
 \be
 \Expectnoc{Q(t)} = \Bra{in} \left[\bar{\rm{T}} \rm{exp} \left(i\int_{t_{i}}^t dt' H(t')\right) \right] Q\left[ \rm{T}\rm{exp}\left(-i \int_{t_{i}}^t dt' H(t')\right)\right]\Ket{in}\,.
 \ee 
We can also calculate correlators using the path integral
\be
 Z[J_+,J_-] = Z_0 \int \D\phi^+\D\phi^- \rm{exp}\left[ i \int_{t_i}^t dt' \int d^3x \left(\LL[\phi^+] -\LL[\phi^-]+J_+\phi^+ + J_-\phi^-\right)\right]\; .
 \ee
Correlation functions are then obtained by varying $Z$ with respect to the sources. Differentiating with respect to $J_+$ brings down a time ordered $\phi$ while differentiating with respect to $J_-$ brings an anti-time ordered $\phi$. 
Because of the doubling of fields, the Green's functions can be arranged in matrix with four entries containing all the different time ordering $G^{\pm\pm}(x,y)$. 
\be
\left(\begin{array}{cc} G^{++}(x,y)& G^{+-}(x,y)\\ G^{-+}(x,y) & G^{++}(x,y)\end{array}\right)\; ,
\ee
\begin{align}
G^{-+}(x,y) & = i\Expectnoc{\phi(x)\phi(y)}  & G^{+-}(x,y)= i\Expectnoc{\phi(y)\phi(x)}\nonumber\; ,\\
G^{++}(x,y)& = i\Expectnoc{\rm{T}\phi(x)\phi(y)}  & G^{--}(x,y) = i\Expectnoc{\bar{\rm{T}}\phi(x)\phi(y)} \; .
\end{align}
They obey the identity $G^{++}(x,y) + G^{--}(x,y) = G^{-+}(x,y) + G^{+-}(x,y)$.
We will use a different basis where
\begin{align}
\phi_C &= \frac{\phi^+ + \phi^-}{2} \; ,& \phi_\Delta = \phi^+ - \phi^-\; ,
\end{align}
and where the Green's function are given by 
\be
 \left(\begin{array}{cc}
  iG_C & G_R \\
  G_A & 0 \\
 \end{array}\right)
 = R \left( \begin{array}{cc}
  G^{++} & G^{+-} \\
  G^{-+} & G^{--} \\
 \end{array} \right) R^T \,,
\ee
for some rotation $R$, with
\bea
G_C(x,y) & = &- \frac{i}{2} \left(G^{-+}(x,y) + G^{+-}(x,y)\right) \nonumber\; ,\\
G_R(x,y)& = & G^{++}(x,y) - G^{+-}(x,y)\nonumber \; ,\\
G_A(x,y) & =& G^{++}(x,y) - G^{-+}(x,y)\; .
\eea
We can build the various propagators from the modes solutions given in \eqref{mf}. It is useful and customary to Fourier transform the space components while leaving the explicit time dependence. 
\bea
G_C(k,\tau_1,\tau_2) & = &  \frac{\pi\sqrt{\tau_1\tau_2}}{4a(\tau_1) a(\tau_2) } \rm{Re}\left( H^{(1)}_{\nu}(-k\tau_1) H_{\nu}^{(1)^*}(-k\tau_2)\right)\; ,\\
G_R(k,\tau_1,\tau_2) & = & -\frac{\pi\sqrt{\tau_1\tau_2}}{2a(\tau_1)a(\tau_2)} \theta(\tau_1 - \tau_2) \rm{Im}\left(H_\nu^{(1)}(-k\tau_1) H_\nu^{(1)^*}(-k\tau_2)\right)\; ,
\eea
with $a(\tau) \approx -1/(H\tau)$ to first order in slow-roll expansion and $\nu=3/2+2\epsilon+\delta$. We will denote the field $\phi_C$ by a solid line and the field $\phi_\Delta$ by a dashed line. 
\begin{figure} [ht]
\begin{center}
\includegraphics[width=0.7\textwidth,angle=0]{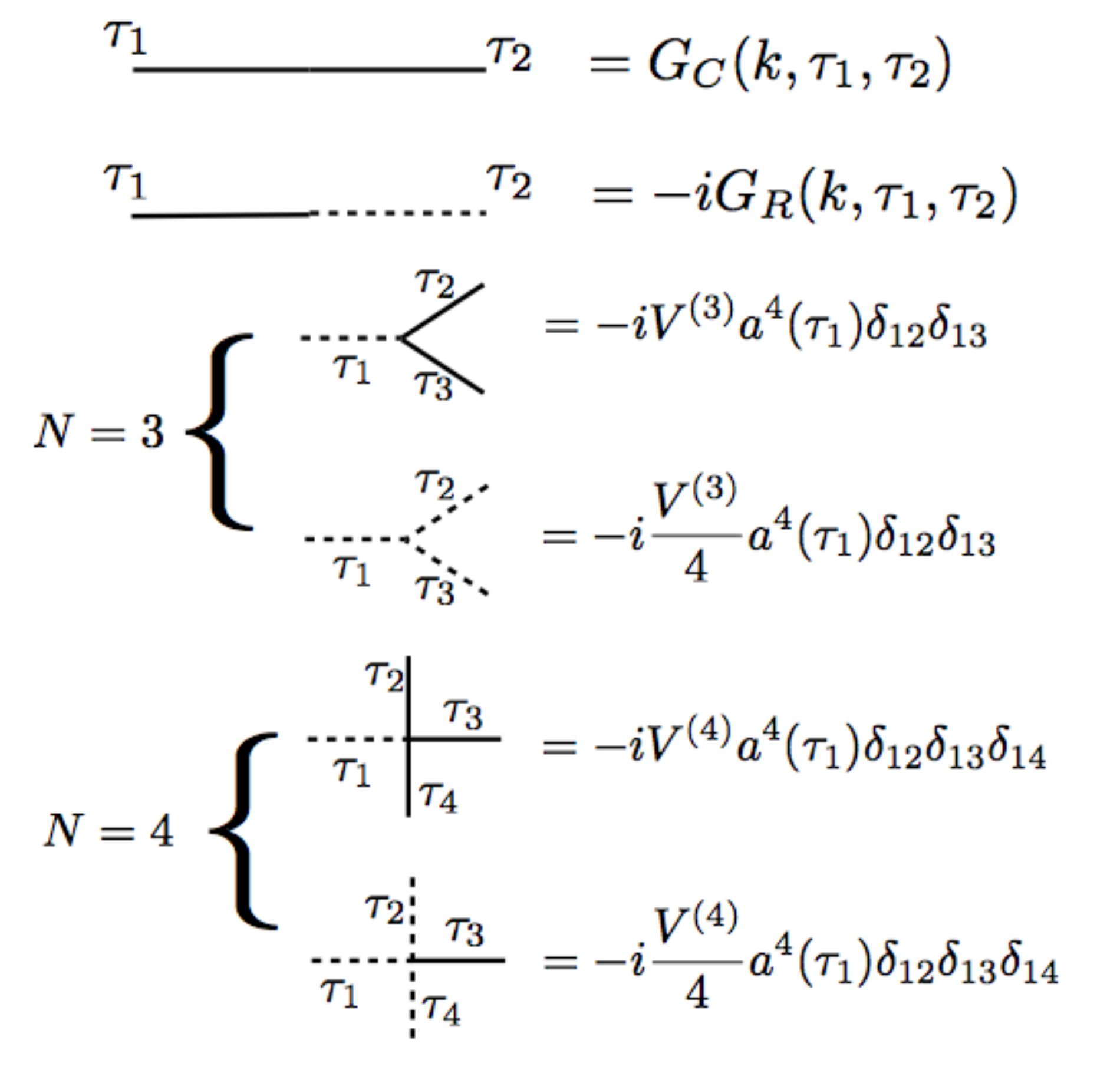}
\caption{Propagators and some vertices ($N=3$ and $N=4$) for this theory. Note that $G_A(k,\tau_1,\tau_2) = G_R(k,\tau_2,\tau_1)$. We use the shorthand notation $\delta_{12} = \delta(\tau_1-\tau_2)$. When a two-point function is attached to a vertex, the corresponding time must be integrated over. Internal spatial momenta must also be integrated over $\int d^3p/(2\pi)^3$. As usual, one must also divide by the symmetry factor for the diagram. At $N= 5$, there are 3 diagrams, etc. }\label{prop}
\end{center}
\end{figure}
To find the Feynman rules we rewrites the potential in terms of $\phi_C$ and $\phi_\Delta$ where for simplicity we denote $\delta\phi \equiv \phi$,
\be
\sum_{N=3}^\infty \frac{V^{(N)}}{N!} ((\phi^+)^N - (\phi^-)^N) = \sum_{N=3}^\infty \frac{V^{(N)}}{N!} \sum_{k=0}^N \left(\begin{array}{c} N\\k\end{array}\right) \frac{1}{2^k} \phi_C^{N-k}\phi_\Delta^k \left(1- (-)^k\right)  \; ,
\ee
where $V^{(N)}$ is given in \eqref{VN}.  In our Feynman rules, the coefficient of a given vertex is the coupling appearing in front of a term
\be
 A^N_{N-k,k} \frac{\phi_C^{N-k} \phi_\Delta^k}{(N-k)! k!}\,,
\ee
which is simply (for $k$ running from $0$ to $N$)
\be
A^N_{N-k,k} = \frac{(1-(-)^k)}{2^k}\; .
\ee
Note that this vanishes for any even k. See figure \ref{prop} for the propagators and the first few explicit vertices. 


 \section{Consistency Relations for Squeezed and Collinear Limits}\label{consistency}
 
In \cite{Maldacena:2002vr} it was realized that, in single field inflationary models, interesting consistency relations arise in the limit where one of the momenta of the three-point correlation function is taken to zero (squeezed or soft limit). As it was later stressed in \cite{Creminelli:2004yq}, the key property is that there exists a single clock on which all correlation functions depend which is valid for all single field models with no assumptions about slow-roll dynamics. The original derivation was done for the bispectrum but the generalization to any $N$-point functions is straightforward \cite{Huang:2006eha,Li:2008gg}. Here we re-derive it in x-space following \cite{Cheung:2007sv}. 

With very analogous reasoning one can also derive a consistency relation for collinear limits, i.e.~when a subset of the momenta of the external legs (and its complement) sum up to zero. This relation has also been used in the literature before \cite{Seery:2008ax,Giddings:2010nc, Gao:2010xk}. We use the notation that prime in $\Expect{}'$ means that the factor $(2\pi)^3\delta^3(\bf K)$ is absent. 


\subsection*{Squeezed (or soft) limit}

Correlation functions in single field inflationary models depend on a single clock and are evaluated at horizon crossing. Using the inflaton as a clock, all $N$-point functions should be evaluated at $\phi = \phi_*$, where $\phi_*$ corresponds to some pivot scale $k_*$ relevant to a specific observation. In a squeezed or collinear limit of a given mode $q\rightarrow 0$, the associated curvature perturbation $\R^B$, where $B$ stands for background, is well outside the horizon, hence it becomes classical and its constant value is determined by the square root of the spectrum. The trick to evaluate an $N$-point function in this limit is to perturb the background by this long wavelength mode $\R_q^B$ as in \eqref{metric}. Assuming a small perturbation $\R^B$, by Taylor expanding one finds
\be\label{Texp}
\Expect{\prod_{i=1}^N\R_{\mathbf{k}_i}}_{\R^B}' =\Expect{ \prod_{i=1}^N \R_{\mathbf{k}_i}  }_{\R_B =0}' + \R_B \left[ \frac{\partial}{\partial\R^B} \Expect{\prod_{i=1}^N\R_{\mathbf{k}_i} }_{\R^B} '\right]_{\R^B=0}  + \mathcal{O}\left(\R_B^2\right)\; .
\ee
A shift in $\R^B$ can be traded for a shift in time, or $\phi_*$ or $k_*$ using $d\R=Hdt=d\phi_*/\sqrt{2\e}=d\log k_*$. If we decide to use $\phi_*$ to  parametrize time, we find 
\be\label{Texp2}
\left.\Expect{ \prod_{i=1}^N\R_{\mathbf{k}_i} }' \right|_{\phi} = \left.\Expect{ \prod_{i=1}^N\R_{\mathbf{k}_i} }' \right|_{\phi_*} + \sqrt{2\e} \R_q \frac{\partial}{\partial\phi_*} \left.\Expect{ \prod_{i=1}^N\R_{\mathbf{k}_i} }'\right|_{\phi_*}  + \cdots\; .
\ee
The squeezed limit is then obtained by multiplying the last expression by some $\R_k$ and taking the expectation value
\be\label{crel}
\lim_{q\rightarrow0} \Expect{\R_{\bf q} \prod_{i=1}^N\R_{\mathbf{k}_i} } = (2\pi)^3\delta^3(\mb{K})\sqrt{2\e} |\R_q|^2  \frac{\partial}{\partial\phi_*} \Expect{ \prod_{i=1}^N\R_{\mathbf{k}_i} }'\; ,
\ee
where it is understood that everything is evaluated at horizon crossing.


\subsection*{Collinear Limit, Scalar and Graviton Exchange}

\begin{figure} [ht]
\begin{center}
\includegraphics[width=0.5\textwidth,origin=c,angle=0]{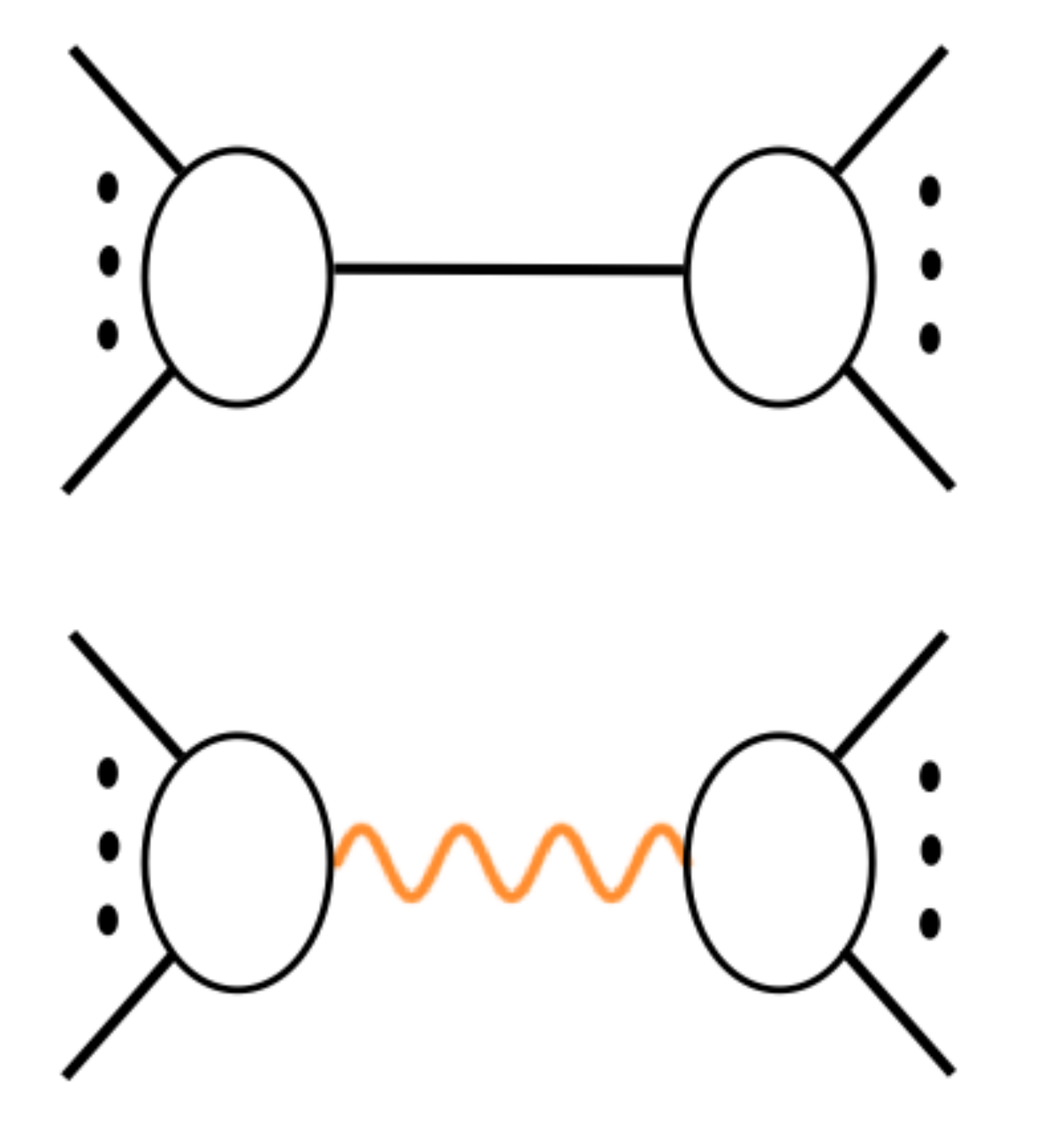}
\caption{The behavior of an $N$-point function in the limit where an intermediate scalar or graviton propagator has vanishing momenta can be obtained using the consistency relations.\label{collinear}}
\end{center}
\end{figure}

A similar consistency relation can be derived in the collinear limit where an intermediate momenta goes to zero, see figure \ref{collinear}. The intermediate state could be a scalar or a tensor and we will discuss both. The first correlation function for which such a limit exists is the trispectrum \cite{Seery:2008ax,Giddings:2010nc}. It is straightforward to find a relation for any $N$-point function.
Consider the case in which a proper subset of the $N$ external momenta, say $1,\cdots, r$, sum up to $q$ and we take the limit $q\rightarrow 0$. The other momenta $r+1, \cdots N$ sum up to zero as well by momentum conservation, such that $\sum_{i=1}^r {\bf k}_i=\sum_{j=r+1}^N {\bf k}_j\equiv {\bf q}$. Then diagrams like the ones in figure \ref{collinear} have an internal propagator going on shell and hence can become very large. This quasi-on-shell propagator that mediates a long-range interaction ($q\rightarrow 0$) can be either a scalar mode or a graviton mode. In the following we will treat each case separately. Let us start considering the scalar-exchange ($SE$) case. We can separately evaluate the correlation function of the first and second subset of momenta in the background of some long wavelength scalar mode $\R^B_q$. For small $\R^B$ we can Taylor expand and use $d\R=d\phi/\sqrt{2\epsilon}$ as in \eqref{Texp2}. 
The consistency relation in the collinear limit is then obtained by multiplying these two expressions and taking the expectation value. The result is
\bea\label{c1}
\lim_{q\rightarrow 0}\Expect{ \prod_{i=1}^N\R_{\mathbf{k}_i} }'^{SE} &= &\Expect{\Expect{\prod_{i=1}^r\R_{\mathbf{k}_i}  }_{\R_q}\Expect{\prod_{i=r+1}^N\R_{\mathbf{k}_i}  }_{\R_{q'}}}' \nn\\
&= &2\e \Expect{\R_{\bf q'}\R_{\bf q}}' \frac{\partial}{\partial\phi_*} \Expect{ \prod_{i=1}^r\R_{\mathbf{k}_i} }' \frac{\partial}{\partial\phi_*} \Expect{ \prod_{i=r+1}^N\R_{\mathbf{k}_i} }'\; .
\eea
Since $|\R_q|^2\propto q^{-3}$, this contribution is divergent in the strict collinear limit at tree level. 

Let us now turn the case in which a long wavelength graviton is exchanged ($GE$). This situation was first analyzed in \cite{Seery:2008ax} for the four-point function. We follow their derivation generalizing it to any $N$-point function and we will work in momentum space. 
Analogously to the case of a scalar mode, we consider a $N$-point correlation function in a background perturbed by a long wavelength classical graviton, which appears in the metric at linear order as
\be \label{mp}
g_{ab}=a^2\left(\delta_{ab}+\gamma_{ab}\right)\,,
\ee 
where $\gamma$ is transverse and traceless. First we want to translate derivatives with respect to $\gamma$ into derivatives with respect to $k_i$. Using the perturbed metric \eqref{mp}, we find 
\be
k^2 \rightarrow k^a k_a + k^a\gamma_{ab} k^b =  k^a k_a - k_a\gamma^{ab} k_b
\ee 
where $a,b=1,2,3$ and the sum over repeated indices is implicit.
Scalar functions, like the $N$-point scalar correlation functions, due to rotational invariance can depend on (spatial) momenta only via combinations like $k_a g^{ab} k_b$. The correlation function $\Expect{\R_\mb{k_1}\cdots\R_{\mathbf{k}_N}}'$ is a function of $\{ \mathbf{k}_1,\dots,\mathbf{k}_{N-1}\}$  and so we can use the chain rule
\be 
\frac{\partial}{\partial \gamma_{ab}}=\sum_{i,j}^{N-1} k^a_ik^b_j\frac{\partial}{\partial ({\bf k}_i\cdot {\bf k}_j)}\,.
\ee
Then, Taylor expanding we find
\be
\Expect{\prod_{i=1}^N\R_{\mathbf{k}_i}  }_{\gamma}' =\Expect{\prod_{i=1}^N \R_{\mathbf{k}_i}  }_{\gamma =0}' + \gamma_{ab}\sum_{i,j} k^a_ik^b_j \left[ \frac{\partial}{\partial ({\bf k}_i\cdot {\bf k}_j)} \Expect{\prod_{i=1}^N\R_{\mathbf{k}_i} }_{\gamma}' \right]_{\gamma=0}  + \mathcal{O}\left(\gamma^2\right)\,.
\ee
For a massless spin-two field like the graviton, the two propagating degrees of freedom can be put together into a causal quantum field as
\bea
\gamma_{ab}({\bf k})&=&\sum_{s=\times,+} \left[\gamma_k(t) b^s_{\bf k}\epsilon^s_{ab}({\bf k})+\gamma_k^*(t) b^{s\dagger}_{-\bf k}\epsilon_{ab}^s(-{\bf k})\right]\,,\nonumber \\
\left[ b^{s}_{{\bf k}} , b^{s' \dagger}_{{\bf k}' } \right] & = & (2\pi)^3 \delta_{s s'} \delta^3({\bf k}-{\bf k'})\,,
\eea
where $\epsilon_{ab}^s({\bf k})$ are the two polarization tensors satisfying $\epsilon_{aa}^s({\bf k})=0
$, $k^a\epsilon_{ab}^s({\bf k})=0$ as well as the completeness relation $\epsilon_{ab}^s({\bf k})\epsilon_{ab}^{s'}({\bf k})=2\delta_{ss'}$. Now, as we did in the scalar case, we can multiply an $N_1$ and $N_2$-point correlation functions in the background of some long wavelength graviton and average over the background. The result written in terms of polarization tensors is
\bea\label{ct}
&&\lim_{q\rightarrow 0}\Expect{ \prod_{i=1}^N\R_{\mathbf{k}_i} }'^{GE} = \Expect{\Expect{\prod_{i=1}^r\R_{\mathbf{k}_i}  }_{\gamma_q}\Expect{\prod_{i=r+1}^N\R_{\mathbf{k}_i}  }_{\gamma_{q'}}}' \\
&& =|\gamma_q|^2 \sum_{ij}^{N_1 -1} \sum_{lm}^{N_2-1}\sum_{s=+,\times} \epsilon^s_{ab}({\bf q})\epsilon^s_{cd}({\bf q})k^a_i k^b_j k^c_l k^d_m  \frac{\partial}{\partial ({\bf k}_i\cdot {\bf k}_j)} \Expect{\prod_{i=1}^{N_1}\R_{\mathbf{k}_i} }'  \frac{\partial}{\partial ({\bf k}_l\cdot {\bf k}_m)} \Expect{\prod_{i=1}^{N_2}\R_{\mathbf{k}_i} }'  \,.\nonumber
\eea 
In order to perform the sum over polarization, we choose an orthonormal basis $\{ {\bf e}^1,{\bf e}^2,{\bf q}/q\}$. 
The polarization tensors can be written as
\be 
\epsilon^{+}_{ab}={\bf e}_a^1{\bf e}_b^1-{\bf e}_a^2{\bf e}_b^2\,,\quad \epsilon^{\times}_{ab}={\bf e}_a^1{\bf e}_b^2+{\bf e}_a^2{\bf e}_b^1\,.
\ee
Using spherical coordinates, each vector can be expressed as 
\be
\mb{k}_i = k_i(\sin\theta_i\cos\phi_i, \sin\theta_i\sin\phi_i,\cos\theta_i)
\ee
where $\cos\theta_i \equiv \mb{k}_i\cdot \mb{q}/(qk_i)$, and $\phi_i$ is the azimuth angle on the plane perpendicular to ${\bf q}$. 
A straightforward computation then shows that
\bea 
\epsilon^+_{ab}({\bf q})k^a_i k^b_j = k_i k_j\sin\theta_i\sin\theta_j\cos(\phi_i+\phi_j)\nonumber \;, \\
\epsilon^{\times}_{ab}({\bf q})k^a_i k^b_j = k_i k_j\sin\theta_i\sin\theta_j\sin(\phi_i+\phi_j)\; . 
\eea
All together, the polarization sum is 
\bea
 E_{ijlm}&\equiv& \sum_{s=+,\times} \epsilon^s_{ab}({\bf q})\epsilon^s_{cd}({\bf q})k^a_i k^b_j k^c_l k^d_m \nonumber\\
 &=& k_ik_j k_lk_m\sin\theta_i\sin\theta_j\sin\theta_l\sin\theta_m \cos(\phi_i +\phi_j-\phi_l-\phi_m)\,.
\eea
Hence we find the general expression
\be\label{cotensor}
\lim_{q\rightarrow 0}\Expect{ \prod_{i=1}^N\R_{\mathbf{k}_i} }'^{GE} = |\gamma_q|^2 \sum_{\{i,j\}=1}^{N_1 -1} \sum_{\{l,m\}=1}^{N_2-1}E_{ijlm}\frac{\partial}{\partial ({\bf k}_i\cdot {\bf k}_j)} \Expect{\prod_{i=1}^{N_1}\R_{\mathbf{k}_i} }'  \frac{\partial}{\partial ({\bf k}_l\cdot {\bf k}_m)} \Expect{\prod_{i=1}^{N_2}\R_{\mathbf{k}_i} }'\; .
\ee
Note that it is understood that the correlation function $\Expect{}'$ have their last momenta eliminated using translation invariance (e.g $k_{N_1} = |\sum_i^{N_1-1} \mb{k}_i|$). This means that even if $\Expect{\R^N}$ is only a function of the norm of each vectors (as we have in \eqref{fullN}), $\Expect{\R^N}'$ depends on angles and the derivatives get quickly very complicated. The special case is $N=4$ with $N_1=N_2 = 2$ where the formula simplifies to 
\be
\lim_{q\rightarrow 0}\Expect{ \prod_{i=1}^4\R_{\mathbf{k}_i} }'^{GE} = |\gamma_q|^2 k_1^2 k_3^2\sin^2\theta_1\sin^2\theta_3 \cos \left[ 2(\phi_1 -\phi_3) \right] 
\frac{\partial}{\partial k_1^2} \Expect{\R_{\mb{k}_1}\R_{-\mb{k}_1} }'  \frac{\partial}{\partial k_3^2} \Expect{\R_{\mb{k}_3} \R_{-\mb{k}_3} }'\; .
\ee
Using the standard definitions of scalar spectral index and tensor-to-scalar ratio
\bea
 n_s-4&\equiv& \frac{\partial \log |\R_k|^2}{\partial \log k} \Big|_{k=k_*}\equiv\frac{\partial \log P_{\R}(k)}{\partial \log k}\Big|_{k=k_*}\,,\nonumber \\
 r & \equiv & \frac{ \Expect{\gamma_{ab} ({\bf k}_*)\gamma^{ab}({\bf k}_*) }' }{\Expect{ \R_{\mb{k}_*}\R_{\mb{k}_*} }'   }  = \frac{4|\gamma_{k_*}|^2}{|\R_{k_*}|^2}\,,\label{r}
 \eea
 for some pivot scale ${\bf k}_*$, we find
 \bea\label{cotensor4}
\lim_{q\rightarrow 0}\Expect{ \prod_{i=1}^4\R_{\mathbf{k}_i} }^{GE} & =& (2\pi)^3\delta^3(\mb{K}) \sin^2\theta_1\sin^2\theta_3 \cos\left[2(\phi_1 -\phi_3)\right]  \frac{r}{4} \left(\frac{n_s-4}{2}\right)^2 P_{\R}(k_1)P_{\R}(k_3)P_{\R}(k_{12})\nn\\
&=&(2\pi)^3\delta^3(\mb{K}) \sin^2\theta_1\sin^2\theta_3 \cos\left[2(\phi_1 -\phi_3)\right] \frac{9r}{16}  P_{\R}(k_1)P_{\R}(k_3)P_{\R}(k_{12})\,,
\eea
where we have use $n_s \approx 1$ in the last line. 


\subsection*{x-space Derivation of the Consistency Relations}

In the following we provide a derivation of the consistency relation \eqref{crel} and \eqref{c1} starting from $x$-space. 
We follow the derivation presented in\footnote{Obtained in collaboration with P.~Creminelli.} \cite{Cheung:2007sv}, generalizing it to any $N$-point functions. Let us start considering the connected $N$-point function $\Expect{\R^N}$ in coordinate space in a quasi de Sitter background slightly perturbed by a small long-wavelength classical mode $\R_B$. By Taylor expanding around $\R_B=0$, we find
\be \label{ini}
\Expect{\R^N }_{\R_B}-\Expect{\R^N }_{0}({\bf x}_1,\dots,{\bf x}_N)=\R_B({\bf x})\left[\frac{\partial}{\partial \R_B}\Expect{\R^N }_{\R_B}({\bf x}_1,\dots,{\bf x}_N)\right]_{\R_B =0}\,.
\ee
Since the background mode $\R_B$ is very slowly changing (in the sense that its momentum ${\bf q}$ is much smaller than the momenta associated with $\R ({\bf x}_i)$ for $i=1,\dots,N$), we can evaluate $\R_B$ at any one of the points $({\bf x}_1,\dots,{\bf x}_N)$ or a linear combination thereof with coefficients of order one. For concreteness and later convenience we choose to evaluate it at ${\bf x}_N$. The derivative with respect to $\R_B$ can be changed into logarithmic derivatives with respect to the coordinates by noticing that an additive shift in $\R$ is equivalent to a multiplicative shift in all coordinates (see e.g.~\eqref{metric}) 
\be 
\frac{\partial}{\partial \R_B}=\sum_{j=1}^N {\bf x}_j\cdot \nabla_j\,,
\ee
where $\nabla_j$ is the gradient with respect to ${\bf x}_j$. Performing a Fourier transform on both sides of \eqref{ini} from $({\bf x}_1,\dots,{\bf x}_N)$ to $({\bf k}_1,\dots,{\bf k}_N)$, we get
\bea
\Expect{\R^N }_{\R_B}-\Expect{R^N }_{0}({\bf k}_1,\dots,{\bf k}_N)&=&\int d^3x_1\dots d^3x_N  e^{-i\sum_j^N {\bf x}_j\cdot{\bf k}_j}\R_B({\bf x}_N)\times\nn\\
&&\times \left[\sum_{j=1}^N {\bf x}_j\cdot \nabla_j\Expect{\R^N }_{\R_B}({\bf x}_1,\dots,{\bf x}_N)\right]_{\R_B =0}\,.\nonumber
\eea
Because of the  translational invariance of the background, the correlation functions do not depend on all of the $3N$ real parameters $({\bf x}_1,\dots,{\bf x}_N)$, but just on $3(N-1)$ of them. One simple way to impose this is to change variables to $\{ \tilde {\bf x}_i,{\bf x}_N \}$ with $\tbx\equiv {\bf x}_i-{\bf x}_N$ for $i=1,\dots,N-1$. Then the correlation functions may be chosen to depend just on $\{ \tilde {\bf x}_i\}$ and not on ${\bf x}_N$. The fact that this coordinate is the same where we chose to evaluate $\R_B$ simplifies considerably the derivation. Changing coordinates and performing the simple $d^3 x_N$ integral, we find
\bea 
\Expect{\R^N }_{\R_B}-\Expect{\R^N }_{0}({\bf k}_1,\dots,{\bf k}_N)&=&\int d^3\tilde x_1\dots d^3\tilde x_{N-1}  e^{-i\sum_j^{N-1} {\bf \tilde x}_j\cdot{\bf k}_j}\R_B({\bf k}_t)\times\nn\\
&&\left[\sum_{j=1}^{N-1} \tilde {\bf x}_j\cdot \nabla_j \Expect{\R^N }'(\tilde {\bf x}_1,\dots,\tilde {\bf x}_{N-1})\right]_{\R_B =0}\,,\nonumber
\eea
where ${\bf k}_t=\sum_i^{N-1} {\bf k}_i$. These identical $3(N-1)$ integrals can be performed using 
\be 
\int dx e^{- i xk} x\frac{\partial}{\partial x}f(x) = -\frac{\partial}{\partial k}\left[k f(k)\right]\,.
\ee
The result is 
\be 
\Expect{\R^N }_{\R_B}-\Expect{\R^N }_{0}({\bf k}_1,\dots,{\bf k}_N)= -\R_B({\bf k}_t)\sum_{j=1}^{N-1}\left[3\Expect{\R^N }'+ {\bf k}_j\cdot \nabla_j \Expect{\R^N }'( {\bf k}_1,\dots, {\bf k}_{N-1})\right]\,,
\ee
where now $\nabla_j$ stands for a divergence with respect to ${\bf k}_j$. If we now multiply both sides by $\R_{\bf q}$ and average over it, we find
\bea 
\Expect{ \R \Expect{\R^N }}  ({\bf q}, {\bf k}_1,\dots,{\bf k}_{N})&=&\lim_{q\rightarrow 0}\Expect{\R_{{\bf q}} \R^N}\\
&=&-\Expect{ \R_{\bf q} \R_{{\bf k}_t} } \sum_{j=1}^{N-1}\left[3\Expect{\R^N }'+ {\bf k}_j\cdot \nabla_j \Expect{\R^N }'\right]\nonumber \\
&=&-(2\pi)^3\delta^3\left({\bf q}+\sum_{i=1}^{N}{\bf k}_i\right) P_{q}\left[3(N-1)\Expect{\R^N }'+\sum_{j=1}^{N-1} {\bf k}_j\cdot \nabla_j \Expect{\R^N }'\right]\,,\nn
\eea
\textit{quod erat demonstrandum}. The collinear relation is also simple to obtain with the correct overall $\delta$ function. 

We can further massage the results by noting that $\sum_{j=1}^{N-1} {\bf k}_j\cdot \nabla_j$ is the dilatation operator.  
Using the fact that the $N$-point function is of degree $-3(N-1)$ and defining a single pivot scale $k_*$ with which we can form dimensionless ratios, we can further simplify the results. We can write (with angular dependence suppressed)  $\Expect{\R^N}' = f(k_i) g\left( \frac{h(k_i)}{k_*}\right)$ where $f$ is a function of degree $-3(N-1)$ and $g$ is a function of degree zero. This means that $h$ is a linear function such that $\sum_{j=1}^{N-1} {\bf k}_j\cdot \nabla_j h(k_i) = h(k_i)$. With this functional form for $\Expect{\R^N}$, the previous expression simplifies to
\bea
\lim_{q\rightarrow 0}\Expect{\R_{{\bf q}} \R^N}&=&(2\pi)^3\delta^3\left({\bf q}+\sum_{i=1}^{N}{\bf k}_i\right) P_{q}\left[ k_* \frac{\partial}{\partial k_*}\Expect{\R^N }'\right]\,,
\eea
which agrees exactly with \eqref{crel}.


\providecommand{\href}[2]{#2}\begingroup\raggedright\endgroup

\end{document}